\documentclass[12pt]{article}
\input axodraw.sty 
\usepackage{epsf}
\usepackage{cite}
\def\beq{\begin{equation}}

\def\beq{\begin{equation}}
\def\eeq{\end{equation}}
\def\beqn{\begin{eqnarray}}
\def\eeqn{\end{eqnarray}}
\def\bea{\begin{eqnarray}}
\def\eea{\end{eqnarray}} 
\def\be{\begin{equation}}
\def\ee{\end{equation}} 

\begin{document}
\begin{titlepage}
\begin{flushright}FSU-981218
\end{flushright} 
\begin{flushright}BNL-HET-98/48
\end{flushright}
\begin{flushright}December, 1998
\end{flushright}
\vspace{2truecm}
\begin{center}
{\large\bf
ASSOCIATED HIGGS PRODUCTION IN THE MSSM
}
\\
\vspace{1in}
{\bf S.~Dawson}\\
{\it Physics Department, Brookhaven National Laboratory,\\
Upton, NY 11973, USA}
\\ 
\vspace{.25in}
{\bf L.~Reina}\\
{\it  Physics Department, Florida State University,\\ 
Tallahassee, FL 32306, USA}
\vspace{1in}  
\end{center}
\begin{abstract} 
  We investigate the inclusive production of the Higgs
  bosons of the minimal supersymmetric model with a heavy
  quark pair, ($t \overline{t}$ and $b {\overline{b}}$), in
  $e^+e^-$ collisions at high energies,
  $\sqrt{s}\!=\!500$~GeV and $\sqrt{s}\!=\!1$~TeV.  The
  ${\cal O}(\alpha_s)$ QCD radiative corrections are
  included and the dependence of the production rates on the
  parameters of the MSSM is explored.  The associated
  production of a Higgs boson with a $b {\overline b}$ pair
  receives large resonant contributions and can have a
  significant rate.
\end{abstract}
\end{titlepage} 
\clearpage

\section{Introduction}
\label{sec:intro}

The search for the Higgs boson is an important objective of present and future
colliders.  In the simplest version of the Standard Model, the Higgs boson is
necessary in order to understand fermion and gauge boson masses.  Once the
Higgs boson has been discovered, it is crucial to measure its couplings to
fermions and gauge bosons.  The couplings to the gauge bosons can be measured
through the associated production processes, $e^+e^-\rightarrow Z\phi$, $q
{\overline q}^\prime\rightarrow W^\pm \phi $, and $q {\overline q}\rightarrow
Z\phi $, and through vector boson fusion, $W^+W^-\rightarrow \phi $ and
$ZZ\rightarrow \phi $, (where $\phi$ is a generic Higgs boson).  The couplings
of the Higgs boson to fermions are more difficult to measure, however.

\begin{figure}[t,b]
\centering
\epsfxsize=4.in
\leavevmode\epsffile{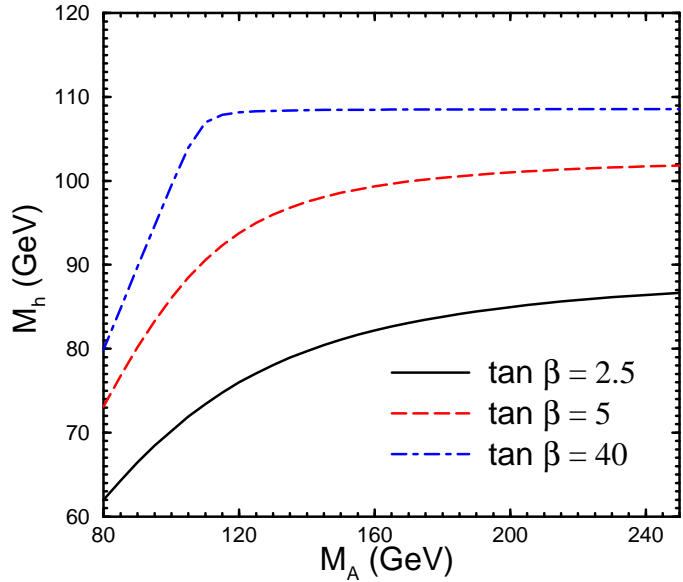}
\caption[]{Mass of the lightest neutral
  Higgs boson as a function of the pseudoscalar mass $M_A$.  The squarks are
  assumed to have a common mass, $M_S\!=\!500$~GeV and the scalar mixing
  parameters are set to zero.}
\label{fig:hmassfig}
\end{figure}

In the Standard Model, the fermion-Higgs couplings are completely determined
in terms of the fermion masses,

\beq
g_{QQh_{SM}}=-{M_Q\over v}\,\,\,,
\label{yuk}
\eeq 

\noindent 
where $v\!=\!(\sqrt{2}G_F)^{-1/2}$ and so the top quark-Higgs boson Yukawa
coupling is large while the bottom quark-Higgs boson Yukawa coupling is much
smaller. In extensions of the Standard Model, however, the Yukawa couplings
can be significantly different.  They are no longer strictly proportional to
the fermion masses, but depend on the parameters of the model.  The minimal
supersymmetric model (MSSM) provides a useful benchmark for comparison of the
$Q{\overline Q}\phi$ couplings with those of the Standard Model.

In fact, the process $e^+e^-\rightarrow t {\overline t} \phi$ provides a
direct mechanism for measuring the top quark-Higgs Yukawa coupling, and for
discriminating the Standar Model type Yukawa coupling from
others.\cite{goun,zer1} This process proceeds via $\gamma$ and $Z$ exchange,
with the Higgs boson bremsstrahlung from the heavy quark.  There is also a
contribution from the Higgs boson coupling to the exchanged $Z$, but this is
always less than a few percent and so does not upset the interpretation of
this process as a measurement of the top quark-Higgs Yukawa couplings.

The dominant production mechanism for a Higgs boson at a hadron machine is
gluon fusion, $gg\rightarrow \phi$, which proceeds by a loop diagram
containing heavy fermions and so is also sensitive to the fermion-Higgs Yukawa
couplings.  For heavy fermions, however, the gluon fusion process is
independent of the fermion mass and so this mechanism counts the number of
heavy fermions which couple to the Higgs boson.  The process $gg\rightarrow
\phi $ thus does not provide an unambiguous determination of the coupling of
the Higgs boson to heavy fermions since it is sensitive to the unknown number
of heavy fermions.
In addition there are uncertainties due to the fact that
QCD corrections to the $gg\rightarrow h$ process are large.   

In a previous paper\cite{dr2}, we computed the ${\cal O}(\alpha_s)$
  corrections to the process $e^+e^-\rightarrow t {\overline t} h_{SM}$.  At
  $\sqrt{s}\!=\!500$~GeV, they are large and positive, while at
  $\sqrt{s}\!=\!1$~TeV, the QCD corrections are small.  In this paper, we
  apply the ${\cal O}(\alpha_s)$ corrections\cite{dr2,ditt} to the production
  of the Higgs bosons of the MSSM and explore the dependence of the production
  rates on the parameters of the MSSM.

It is more difficult to extract the bottom quark-Higgs Yukawa coupling from a
measurement of $e^+e^-\rightarrow b {\overline b} h_{SM}$, since the coupling
itself is tiny and the $Z$ contribution is relevant.  However in the MSSM, for
certain values of $\tan\beta$, the ratio of the neutral Higgs vacuum
expectation values, the $b{\overline b}\phi$ Yukawa coupling receives
significant enhancements and so the process $e^+e^-\rightarrow b {\overline
b}\phi$ may be   larger than in the 
Standard Model.    Moreover, there are $5$ Higgs bosons in the MSSM,
so that additional processes not present in the Standard Model may be useful
to pin down the fermion-Higgs boson Yukawa couplings.
In particular, there are  new resonance contributions, for example
 $e^+e^-\rightarrow A^0h_i, A^0\rightarrow b {\overline b}$, which are not
present in the Standard Model.  The effects of these new contributions
can be significant and are the focus of this paper.

The process $e^+e^-\rightarrow b {\overline b} \phi $ is not only sensitive to
the b-quark Higgs Yukawa coupling but also has an interesting dependence on
the other parameters of the supersymmetric model.  Sections 5 and 6 explore
this dependence in detail.

In Section~\ref{sec:susybck}, we provide an overview of the elements of the
MSSM relevant for our study.  The following sections describe our results for
the associated production of the scalars and pseudoscalar of the MSSM with $t
{\overline t}$ and $b {\overline b}$ pairs.  Section~\ref{sec:concl} contains
some conclusions.

\section{SUSY Background}
\label{sec:susybck}

The minimal supersymmetric version of the standard model
(MSSM) is a theoretically appealing extension of the
Standard Model and provides a useful framework for
comparison with the Standard Model.  In the MSSM, the Higgs
sector contains two $SU(2)_L$ doublets and so there are five
physical Higgs bosons; two neutral scalars, $h^0$ and $H^0$,
(which we will often denote by $h_i^0$), a pseudo-scalar
$A^0$, and a pair of charged scalars, $H^\pm$.\cite{hhg} We
are therefore led to consider the processes,

\beqn
e^+e^-&\rightarrow & Q {\overline Q} h^0\,\,\,,
\nonumber  \\ 
e^+e^-&\rightarrow & Q {\overline Q} H^0\,\,\,,\nonumber    \\ 
e^+e^-&\rightarrow & Q {\overline Q} A^0\,\,\,,\nonumber    \\ 
e^+e^-&\rightarrow & {\overline t} b H^+,~~ t {\overline b} H^-\,\,\,,
\nonumber  \\ 
\eeqn

\noindent
where $Q$ is either $t$ or $b$.  These processes have been
previously considered in Ref. \cite{zer1} and we extend
their analysis.  The rates are small, but the signatures are
distinctive.\footnote{The processes $e^+e^-\rightarrow
  {\overline t} b H^+,~~ t {\overline b} H^-$ are 
  interesting in the small mass region where $H^\pm$
  cannot decay to $t$ nor $t$ decay to
  $H^\pm$\cite{zer1}.  Outside this mass region, other processes
	are more useful for determining the Yukawa couplings.\cite{zer1} 
 We will not consider these processes
  further.}

At lowest order, all masses and couplings in the Higgs
sector can be computed in terms of two parameters, which are
typically taken to be $M_A$, the mass of the pseudoscalar
Higgs boson, and $\tan \beta$, the ratio of vacuum
expectation values.  At higher order, there is a dependence
on the scalar masses and mixing parameters.  Having picked
$M_A$, $\tan\beta$, and the scalar masses and mixing
parameters, the neutral Higgs boson masses are
predicted\cite{spir,lepstud,car}.

%The mixing in the scalar sector is defined by the mass matrix-squared
%of the scalar partners of the right and left handed top quarks,
%\beq
%{\tilde M}^2=\left(
%\begin{array}{ll}
%{\tilde M}_Q^2+M_t^2& M_t(A_t-\mu\cot\beta)\\
%~~~~+M_Z^2({1\over 2}-{2\over 3}
%\sin^2\theta_W)\cos 2\beta & ~~~\\
%M_t(A_t-\mu\cot\beta)&{\tilde M}_t^2+M_t^2
%\\
%~~&~~~~+{2\over 3}M_Z^2\sin^2\theta_W\cos 2\beta
%\end{array}\right)
%\quad,
%\eeq
%where ${\tilde M}_Q$ is the mass of the $SU(2)$ doublet of the
%scalar partners of $t_L$ and $b_L$ and ${\tilde M}_t$ is the mass
%of the scalar partner of $t_R$.  We will assume ${\tilde M}_Q={\tilde M}_t
%\equiv M_S$.  We will further assume that the mixing parameters are
%the same in the top and bottom squark sectors, $A_t=A_b=A$.
 
The MSSM has the feature that there is an upper limit on the
mass of the lightest neutral Higgs boson. Including the
radiative corrections to leading logarithm, we have,

\beq 
M_h^2 < M_Z^2 \cos^2 2 \beta +\delta\,\,\,,
\eeq 

\noindent
with 

\beq \delta= {3 G_F  M_t^4\over  \sqrt{2}\pi^2}
 \log\biggl(1+{M_S^2\over M_t^2}\biggr) , 
\eeq 

\noindent and $M_S$ is a common squark mass.  In our numerical studies
the renormalization group improved values for the Higgs
boson masses, along with the NLO corrections are included,
using the program HDECAY.\cite{hdecayref}
Fig.~\ref{fig:hmassfig} shows the mass of the lightest
neutral Higgs boson as a function of $M_A$.  There is
clearly a maximum value around $M_h\!=\!110$~GeV.  Including
mixing effects in the squark and Higgs sectors
% (non-zero $A$ and $\mu$)
raises this limit to around $130$~GeV.  Such a light Higgs
boson may be observable in association with a heavy quark
pair, providing additional impetus to our study.

The LEP2 experiments have searched for the processes $e^+e^-
\rightarrow Zh^0$ and $e^+e^-\rightarrow h^0A^0$ and find the
$95\%$ confidence level limits,\cite{carl}

\beqn
M_A & > & 78\,\,\mbox{GeV}\,\,\,,\nonumber \\
\tan\beta &>&  2.1\,\,\,, 
\label{explims}
\eeqn
\noindent
for the case with no mixing in the squark
sector.\footnote{There is also a small allowed region with
  $\tan\beta\!<\!0.8$.\cite{carl}} We will therefore
consider $\tan\beta\!=\!2.5,5$ and $40$ and $M_A\!
>\!80$~GeV\footnote{ Results from precision electroweak
  measurements suggest that the true bound on $M_A$ may be
  slighter than this.\cite{dp}.}  as representative values in
our study.\footnote{Of course,
for a given value of $\tan\beta$, the lower bound on
$M_A$ may be greater than that of Eq. \ref{explims}.}
   The corresponding values for the light Higgs
mass, $M_h$, can be found from Fig.~\ref{fig:hmassfig}.

\begin{table}[bt]
\begin{center}
{Table 1: Higgs boson  couplings to fermions in units of
  $g_{QQh_{SM}}\equiv -M_Q/v$}\vskip6pt
\renewcommand\arraystretch{1.2}
\begin{tabular}{|lccc|}
\hline
\multicolumn{1}{|c}{$Q$}& $C_{QQh}$& $C_{QQH}$
 & $C_{QQA}  $
\\
\hline
$t$   &    ${\cos\alpha\over \sin\beta}$ &
    ${\sin\alpha\over\sin\beta}$
& $\cot\beta$ \\ 
$b$   &    $-{\sin\alpha\over\cos\beta}$ & 
     ${\cos\alpha\over\cos\beta}$ 
& $\tan\beta$  \\
\hline
\end{tabular}
\end{center}
\end{table}

An important feature of SUSY models is that the
fermion-Higgs couplings are no longer strictly proportional
to mass as they are in the Standard Model.  It is convenient
to write the couplings for the neutral Higgs bosons to the
fermions in terms of the Standard Model Higgs boson
couplings. In particular, we write the couplings of the
Higgs bosons to the quarks as

\beq
{\cal L}=-(\sqrt{2}G_F)^{1/2}M_i \biggl[C_{QQh}{\overline Q}_i Q_i h^0
+C_{QQH} {\overline Q}_i Q_i H^0
+C_{QQA}{\overline Q}_i \gamma_5 Q_i A^0\biggr],  
\eeq

\noindent
where $C_{QQh}$ is $1$ for a Standard Model Higgs boson, 
($C_{QQH}\!=\!C_{QQA}\!=\!0$), while the corresponding MSSM
couplings are given in Table 1.  The fermion-Higgs boson
Yukawa couplings in the MSSM are then,

\beqn 
g_{QQh_i}&=&- (\sqrt{2}G_F)^{1/2}M_Q C_{QQh_i}\,\,\,,
\nonumber \\
g_{QQA}&=& -(\sqrt{2}G_F)^{1/2}M_Q C_{QQA}\,\,\,. 
\label{yukdef} 
\eeqn

\noindent
For small $M_A$, the couplings of the neutral Higgs boson to
fermions can be significantly different from the Standard
Model couplings; the $b$-quark coupling to the lighter of
the neutral Higgs bosons, $h^0$, becomes enhanced, while the
$t$-quark coupling to $h^0$ is suppressed for large
$\tan\beta$.  On the other hand, the coupling of the heavier
neutral Higgs boson, $H^0$, to the $b$ is enhanced at small
$\tan\beta$ and the coupling to $t$ enhanced at large
$\tan\beta$ for light $M_A$.

The angle $\alpha$ is related to the mixing in the neutral
Higgs boson mass matrix.  In the leading logarithmic
approximation,

\beq
\tan 2\alpha={(M_A^2+M_Z^2) \sin 2\beta
\over (M_A^2-M_Z^2)cos 2 \beta +\delta/\sin^2\beta }\,\,\,,
\eeq

\noindent
where $-\pi/2 \!<\! \alpha \!<\! 0$.  In our numerical studies we
include the renormalization group improved values for
$\alpha$ using the program HDECAY.\cite{hdecayref}

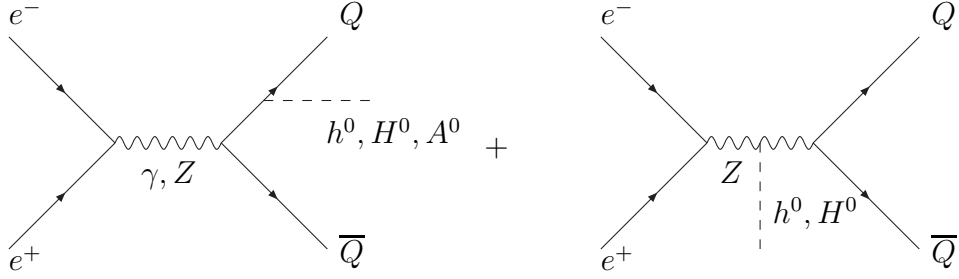
\begin{figure}[bt]
\begin{picture}(100,100)(-20,-5)
\SetScale{0.8}
\ArrowLine(0,100)(50,50)
\ArrowLine(0,0)(50,50)
\Photon(50,50)(100,50){3}{6}
\ArrowLine(100,50)(150,100) 
\ArrowLine(100,50)(150,0)
\DashLine(120,70)(170,70){5}
\put(180,35){$+$} 
\put(50,25){$\gamma, Z$} 
\put(0,85){$e^-$}
\put(0,-8){$e^+$}
\put(120,40){$h^0,H^0,A^0$}
\put(125,85){$Q$}
\put(125,-5){${\overline Q}$}
\SetScale{1}
\end{picture}
\begin{picture}(100,100)(-140,-5)
\SetScale{0.8}
\ArrowLine(0,100)(50,50)
\ArrowLine(0,0)(50,50)
\Photon(50,50)(100,50){3}{6}
\DashLine(75,50)(75,0){5}
\ArrowLine(100,50)(150,100) 
\ArrowLine(100,50)(150,0)
\put(45,25){$Z$} 
\put(0,85){$e^-$}
\put(0,-8){$e^+$}
\put(65,10){$h^0,H^0$}
\put(125,85){$Q$}
\put(125,-5){${\overline Q}$}
\SetScale{1}
\end{picture} 
\vskip .2in 
\caption[]{Feynman diagrams contributing to the lowest
  order processes, $e^+e^-\rightarrow Q {\overline Q} h_i^0,A^0$
  in the MSSM.}
\label{fig:lofeyndiag}
\end{figure}

\begin{figure}[bt]
\begin{picture}(100,100)(-20,-5)
\SetScale{0.8}
\ArrowLine(0,100)(50,50)
\ArrowLine(50,50)(0,0)
\Photon(50,50)(100,50){3}{6}
\DashLine(100,50)(150,100){5}  
\DashLine(100,50)(150,0){5} 
\ArrowLine(150,0)(200,15)
\ArrowLine(200,-15)(150,0) 
\put(54,20){$Z$} 
\put(105,20){$A^0$} 
\put(0,85){$e^-$}
\put(0,-5){$e^+$}
\put(125,85){$h^0,H^0$}
\put(175,-15){${\overline Q}$}
\put(175,15){$Q$} 
\SetScale{1}
\end{picture}
\begin{picture}(100,100)(-140,-5)
\SetScale{0.8}
\ArrowLine(0,100)(50,50)
\ArrowLine(50,50)(0,0)
\Photon(50,50)(100,50){3}{6}
\DashLine(100,50)(150,100){5}  
\DashLine(100,50)(150,0){5} 
\ArrowLine(150,0)(200,15)
\ArrowLine(200,-15)(150,0) 
\put(54,20){$Z$} 
\put(105,20){$h^0,H^0$} 
\put(0,85){$e^-$}
\put(0,-5){$e^+$}
\put(125,85){$A^0$}
\put(175,-15){${\overline Q}$}
\put(175,15){$Q$} 
\SetScale{1}
\end{picture}
\vskip .2in 
\caption{Feynman diagrams contributing to 
  $e^+e^-\rightarrow Q {\overline Q} h_i^0,A^0$ which are
not present in
  the Standard Model.}
\label{fig:ahfig}
\end{figure}
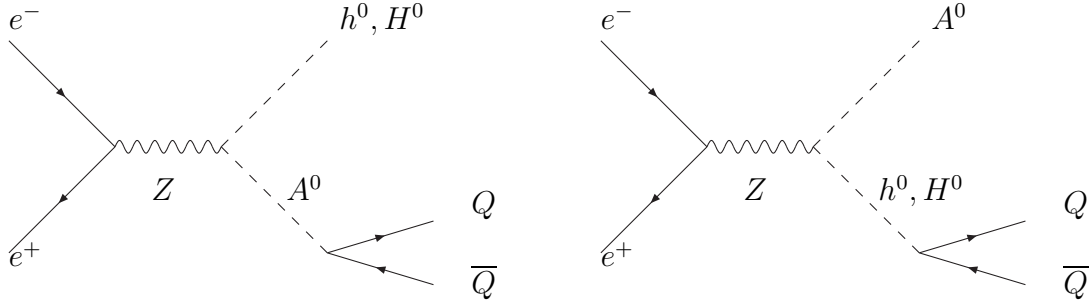

\section{$e^+e^-\rightarrow t {\overline t} h_i^0$}
\label{sec:tth}

The rate for $e^+e^-\rightarrow t {\overline t} h_i^0$,
$h_i^0\!=\!h^0,H^0$, receives the contributions from
$\gamma$ and $Z$ exchange shown in Fig.~\ref{fig:lofeyndiag}
as in the Standard Model.  In the MSSM, there is an
additional contribution not present in the Standard Model
from the sub-process $e^+e^-\rightarrow A^0h^0_i,
A^0\rightarrow t {\overline t}$, shown in
Fig.~\ref{fig:ahfig}.

The rate for the associated production of a neutral SUSY
Higgs boson with a heavy quark can be written
as\cite{zer1,dr2}

\bea 
\frac{d\sigma(e^+e^-\rightarrow Q\bar Q h_i^0)}{dx_h} &=& 
N_c\frac{\sigma_0}{(4\pi)^2}\left\{
\left[Q_e^2Q_Q^2+\frac{2Q_eQ_Qv_ev_Q}{1-M_Z^2/s}+
\frac{(v_e^2+a_e^2)(v_Q^2+a_Q^2)}{(1-M_Z^2/s)^2}\right]G_1\right.
\nonumber\\
&&\left.+ \frac{v_e^2+a_e^2}{(1-M_Z^2/s)^2}\left[a_Q^2\sum_{i=2}^6G_i+
v_Q^2(G_4+G_6)
 +{G_7\over 16 c_W^2 s_W^2}\right]   
\right. 
\nonumber \\
&& \left.  +
\frac{Q_eQ_Qv_ev_Q}{1-M_Z^2/s}G_6 \right\}
\,\,\,\,,
\label{dsig0}
\eea

\noindent where 
$\sigma_0\!\!=\!\!4\pi\alpha^2/3s$, $\alpha$ is the QED fine
structure constant, $N_c\!=\!3$ is the number of colors,
$x_h\!=\!2 E_h/\sqrt{s}$ with $E_h$ the energy of the Higgs
boson $h_i^0$, and $Q_i$, $v_i$ and $a_i$ ($i\!=\!e$, $Q$)
denote the electromagnetic and weak couplings of the
electron and of the heavy quark respectively,

\be 
v_i=\frac{2I_{3L}^i-4Q_is_W^2}{4s_Wc_W}\,\,\,\,\,\,\,\,,\,\,\,\,\,\,\ 
a_i= \frac{2 I_{3L}^i}{4s_Wc_W}\,\,\,\,, 
\ee

\noindent with $I_{3L}^i\!=\!\pm 1/2$ being the weak isospin of the
left-handed fermions and $s_W^2\!=\!1-c_W^2=0.23$.  The
contribution from the $ZA^0h_i^0$ coupling shown in
Fig.~\ref{fig:ahfig} and its interference with the other
contributions is given by the factor $G_7$.  For
completeness we list the coefficients $G_i(x_h)$ in
Appendix~\ref{app:gcoeff}.

\begin{figure}[t,b]
\centering
\epsfxsize=4.in
\leavevmode\epsffile{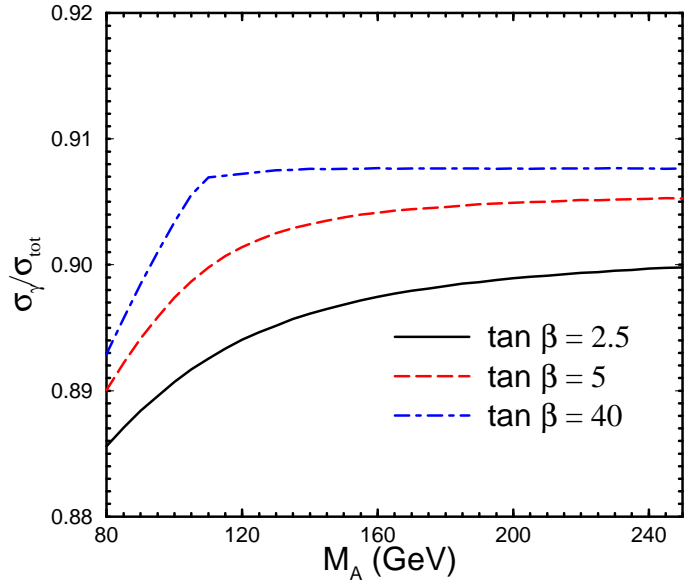}
\caption[]{Dominance of the photon contribution to the
  process $e^+e^-\rightarrow t {\overline t} h^0$ at
  $\sqrt{s}\!=\!500$~GeV.
The squarks are assumed to have a common mass,
  $M_S\!=\!500$~GeV and the scalar mixing parameters are set
  to zero.}
\label{fig:photfig}
\end{figure}

In the case of $t{\overline t}h^0$ and $t{\overline t}H^0$
production, the dominant contribution at
$\sqrt{s}\!=\!500$~GeV is photon exchange, as is the case
also in the Standard Model.  In Fig.~\ref{fig:photfig}, we
show the ratio of the cross section computed including only
the photon exchange to the total cross section for the case
of $e^+e^-\rightarrow t\bar t h^0$. Very similar results hold
for the $e^+e^-\rightarrow t\bar t H^0$ process.
\noindent
Since the cross section is dominated by photon exchange, the
contributions from $e^+e^-\rightarrow A^0h_i^0\rightarrow t
{\overline t} h_i^0$ and $e^+e^-\rightarrow Z\rightarrow t
{\overline t}h^0_i$ can safely be neglected at
$\sqrt{s}\!=\!500$~GeV.

\begin{figure}[t,b]
\centering
\epsfxsize=4.in
\leavevmode\epsffile{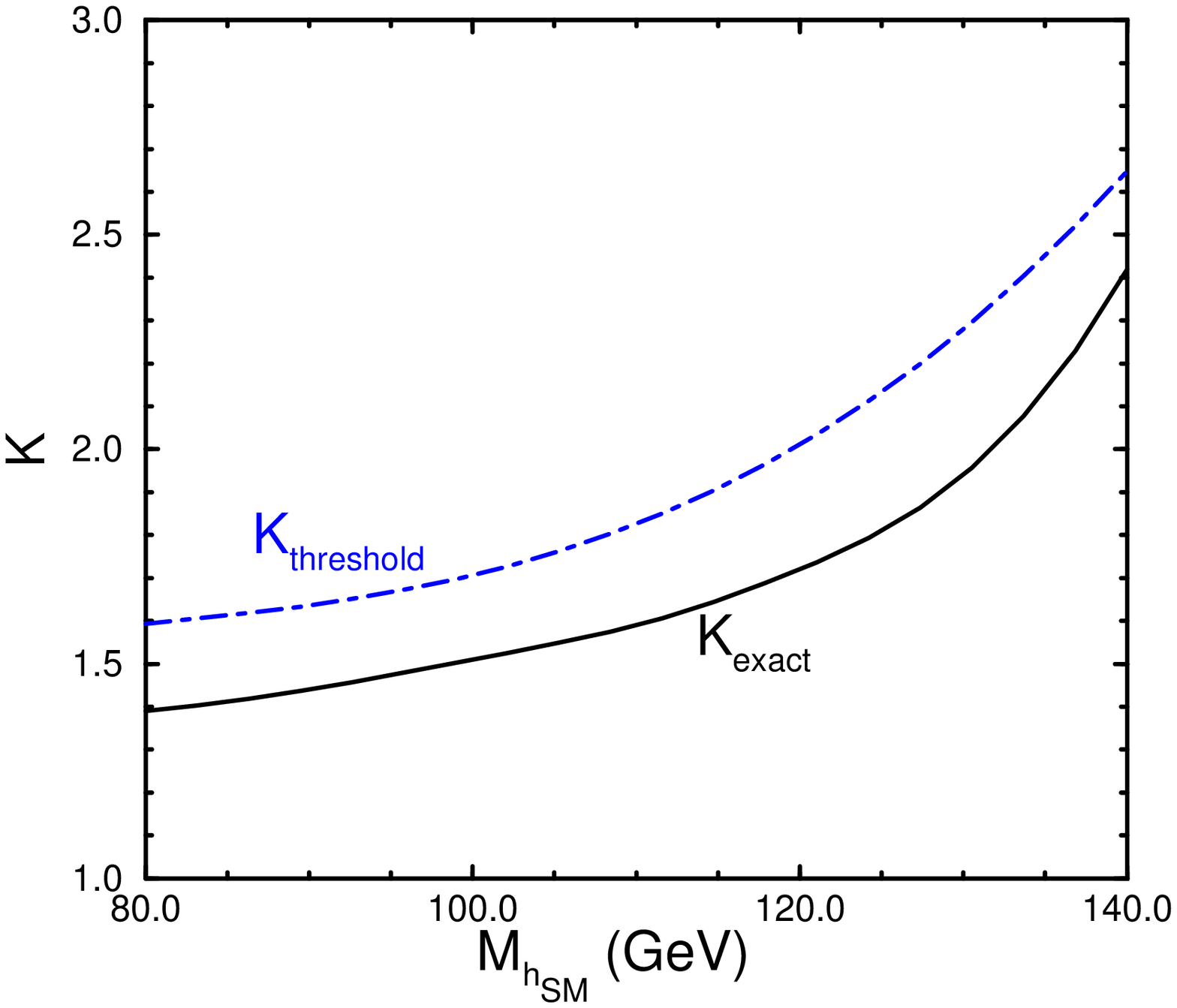}
\caption[]{$K$ factor for the
  process $e^+e^-\rightarrow t {\overline t} h_{SM}$ at
  $\sqrt{s}\!=\!500$~GeV.  This figure uses
  $\alpha_s(M_t)=0.11164$.  $K_{exact}$ is the result of Ref.
  \cite{dr2} and $K_{threshold}$ includes only the $t
  {\overline t}$ threshold contribution.\cite{ditt}}
\label{fig:kfacfig}
\end{figure} 

The ${\cal O}(\alpha_s)$ radiative corrections can be
summarized by a $K$ factor multiplying the lowest order
rate.  The $K$ factor can be defined as 

\beq 
K\equiv{\sigma_{NLO}\over \sigma_{LO}} \quad .  
\eeq

From Fig.~\ref{fig:photfig}, it is clear that the $K$ factor
computed in Ref. \cite{dr2}, where only the $\gamma$
exchange was considered, can be used at
$\sqrt{s}\!=\!500$~GeV.  The $K$ factor depends only on the
mass of the produced Higgs particle, and not on the SUSY
parameters in the limit in which only the photon exchange
contribution is important and so the results for the
Standard Model Higgs boson can be used to obtain the ${\cal
  O}(\alpha_s)$ corrections.  The only dependence on the
renormalization scale $\mu$ is in the strong coupling
constant, $\alpha_s(\mu)$. We will take
$\mu\!=\!M_t\!=\!175$~GeV everywhere.  A useful numerical
fit to the $K$ factor of Ref. \cite{dr2}, valid for
$\sqrt{s}\!=\!500$~GeV, is,

\begin{equation}
K\mid_{\sqrt{s}=500\,\mbox{GeV}}
\sim 1+\alpha_s(\mu^2)\biggl\{
{64\over 9}{M_t\over \sqrt{(\sqrt{s}-M_h)^2-
4M_t^2}}+a_0+a_1r+a_2 r^2\biggr\}\,\,\,,
\label{kfacfit}
\end{equation}
where 
\begin{equation}
r\equiv{M_h\over 100\,\,\mbox{GeV}}\,\,\,,
\end{equation}
and
\begin{eqnarray}
a_0&=& -1.760\,\,\,,
\nonumber \\
a_1&=&-.004\,\,\,,
\nonumber \\
a_2&=& -.133\,\,\,.
\end{eqnarray}

\noindent 
This fit (valid only at $\sqrt{s}\!=\!500$~GeV), is an
excellent approximation to the exact $K$ factor of Refs.
\cite{dr2,ditt}.  Fig.~\ref{fig:kfacfig} shows that the QCD
corrections increase the rate significantly at
$\sqrt{s}\!=\!500$~GeV.  At $\sqrt{s}\!\sim\! 500$~GeV, much
(but not all) of the $K$ factor can be accounted for by the
threshold contribution to $t {\overline t}$ production.  The
first term in Eq.~(\ref{kfacfit}) is the threshold
contribution to the $K$ factor\cite{ditt}, which is plotted as the
curve $K_{threshold}$ in Fig.~\ref{fig:kfacfig},
and the rising of the $K$ factor signaling the Coulomb
singularity as the threshold is approached, $\sqrt{s}\sim 2
M_t+M_{h_{SM}}$, is clearly seen.

\begin{figure}[t,b] 
\centering
\epsfxsize=4.in 
\leavevmode\epsffile{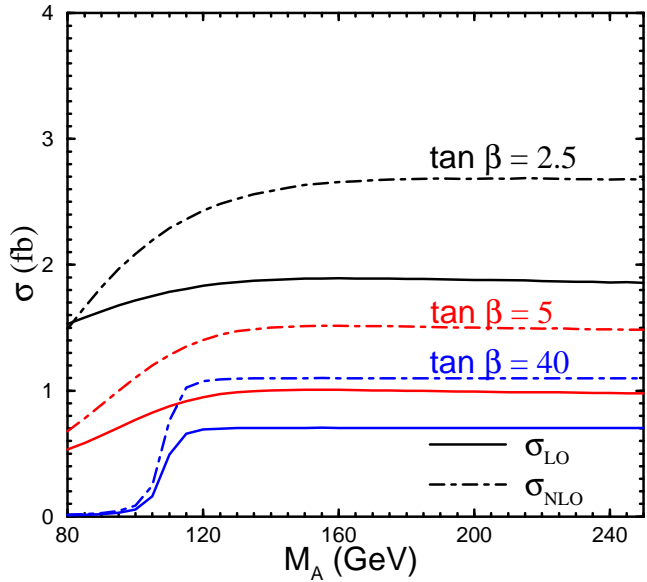}
\caption[]{Cross section for $e^+e^-\rightarrow t {\overline t}h^0$
  production  at $\sqrt{s}=500~GeV$
at lowest order and at NLO for different values
  of $\tan\beta$. We take $\alpha_s(M_t)\!=\!0.11164$ and
  use the $K$ factor of Ref. \cite{dr2}.  The squarks are
  taken to have a common mass, $M_S\!=\!500$~GeV, and the mixing parameters
are set to zero.}
\label{fig:neutfig1}
\end{figure}

\begin{figure}[t,b]
\centering
\epsfxsize=4.in
\leavevmode\epsffile{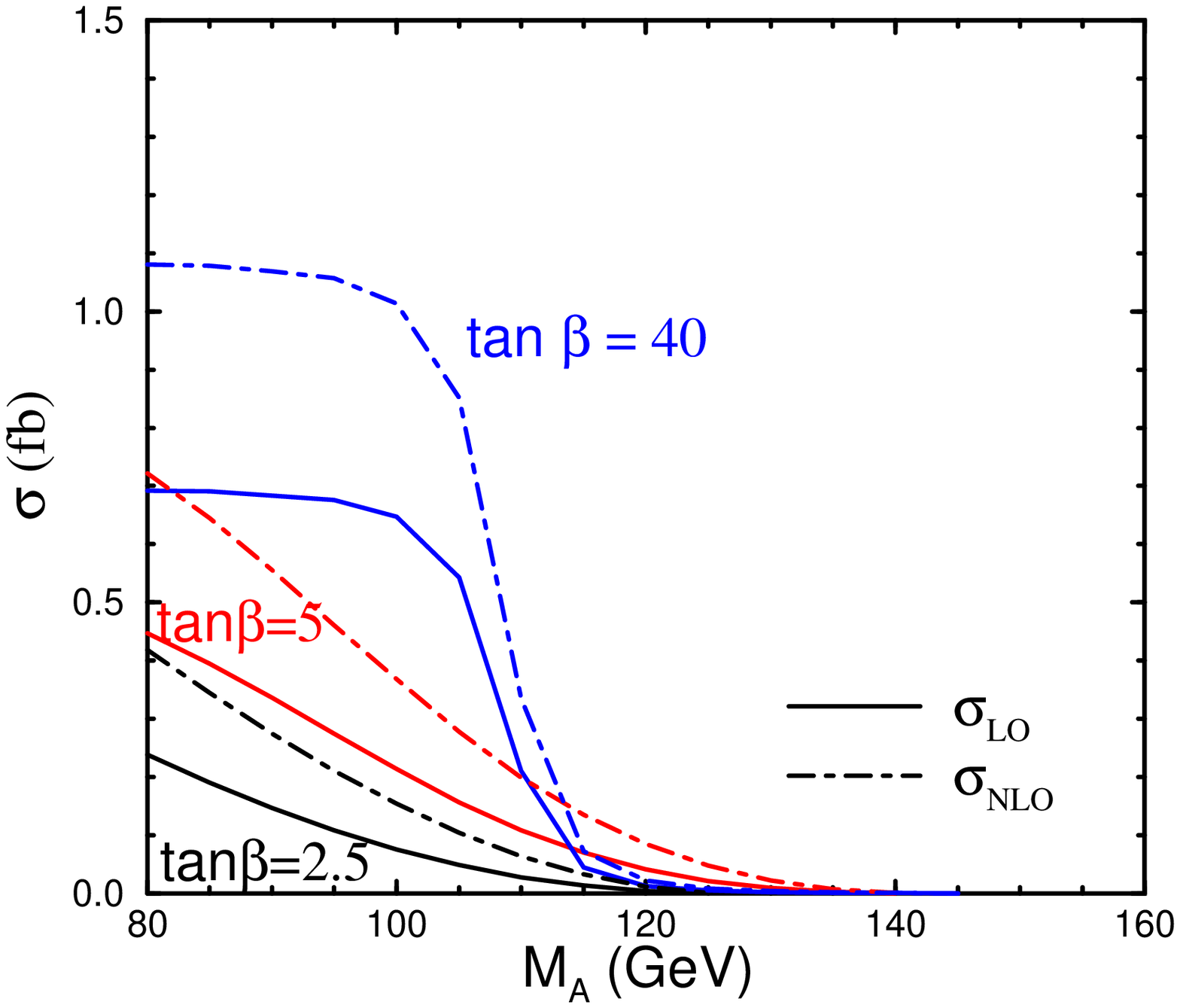}
\caption[]{Cross section for $e^+e^-\rightarrow t {\overline t}H^0$
  production  at $\sqrt{s}=500~GeV$ 
at lowest order and at NLO for different values
  of $\tan\beta$.  We take $\alpha_s(M_t)\!=\!0.11164$ and
  use the $K$ factor of Ref. \cite{dr2}.  The squarks are
  taken to have a common mass, $M_S\!=\!500$~GeV and the
  mixing parameters are set to zero.}
\label{fig:neutfig2}
\end{figure}

We can now compute the rates for $e^+e^-\rightarrow t
{\overline t} h_i^0$ production including the ${\cal
  O}(\alpha_s)$ corrections.  In Figs.~\ref{fig:neutfig1}
and \ref{fig:neutfig2}, we show the lowest order and the
radiatively corrected rates.  All contributions to the
lowest order rate ($Z$ exchange, $A^0 h_i^0$ production, etc)
are included and the next-to-leading order result is
computed by multiplying the lowest order rate by the $K$
factor of Fig.~\ref{fig:kfacfig}.  Since the dominant
contribution is photon exchange, this should be an excellent
approximation.  The rate for $t{\overline t}h^0$ is enhanced
at small $\tan\beta$, while the rate for $t{\overline t}H^0$
production is enhanced at large $\tan\beta$.  The rate is
never larger than a few ${\it fb}$, making this process
extremely challenging to observe.


\begin{figure}[t,b]
\centering
\epsfxsize=4.in
\leavevmode\epsffile{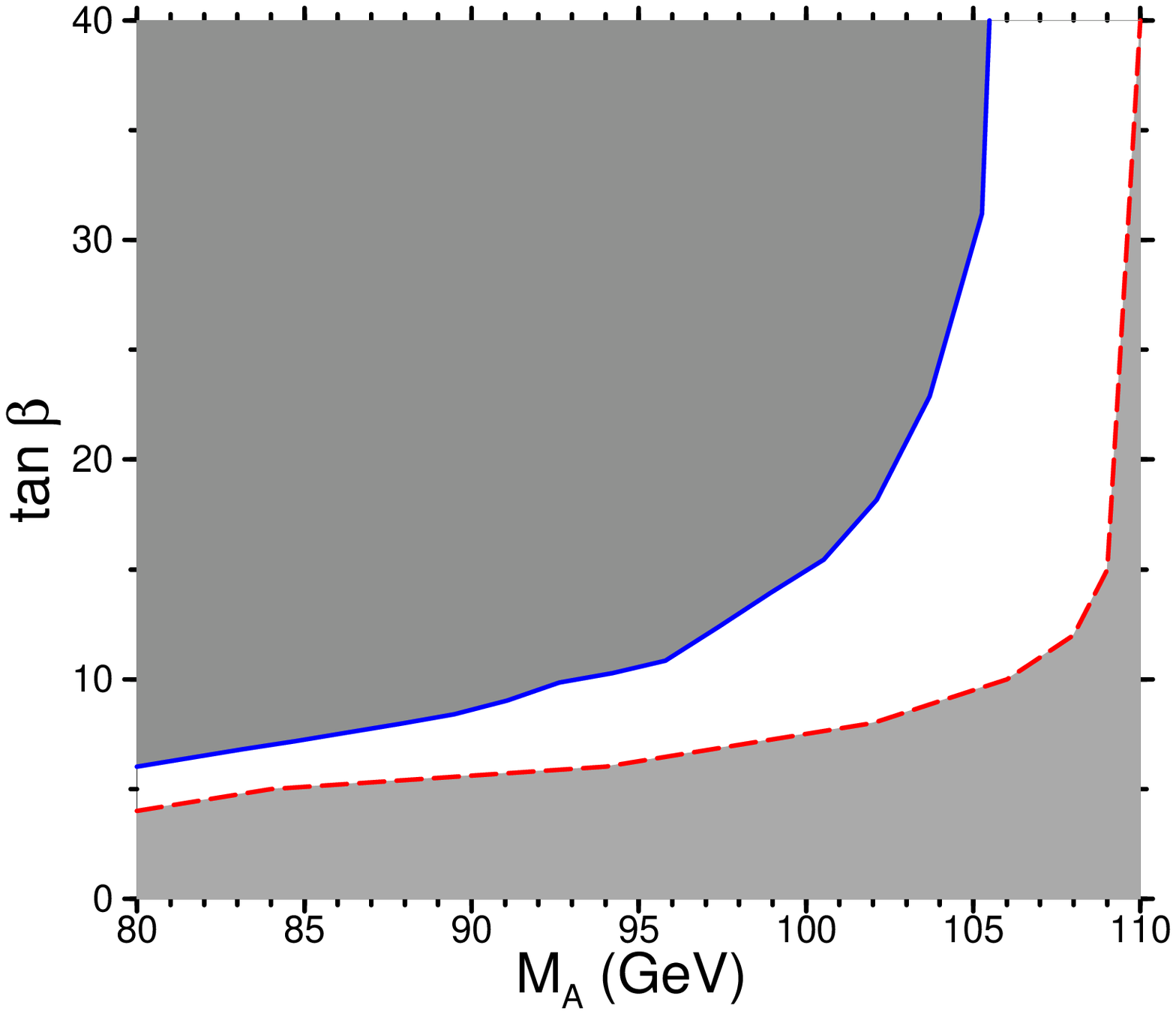}
\caption[]{Regions in the $M_A-\tan\beta$ plane where the
  cross section for $e^+e^-\rightarrow t {\overline t} h_i^0$,
  ($h_i^0\!=\!h^0,H^0$), production is larger than $0.75$~fb at
  $\sqrt{s}\!=\!500$~GeV.  The upper left hand region results from
  $e^+e^-\rightarrow t {\overline t} H^0$, while the region at the lower right
  is the result from $e^+e^-\rightarrow t {\overline t} h^0$.  All NLO QCD
  corrections are included. The squarks are taken to have a common mass,
  $M_S=500~GeV$ and the mixing parameters are set to zero.}
\label{fig:scanfig}
\end{figure}

Although the processes $e^+e^-\rightarrow t {\overline t}h^0$
and $e^+e^-\rightarrow t {\overline t}H^0$ have small rates,
the signatures are distinctive since the final state will be
predominantly $W^+W^-b {\overline b} b {\overline b}$ and so
may be observable with a small number of events.  In
Fig.~\ref{fig:scanfig}, we show the region of the
$M_A-\tan\beta$ plane where the rates for either process are
larger than $0.75$~fb and we see that this includes much of
the parameter space.  We have everywhere set the mixing in
the squark and Higgs sectors to zero.  Our results are
relatively insensitive to this assumption.

\begin{figure}[t,b]
\centering
\epsfxsize=4.in
\leavevmode\epsffile{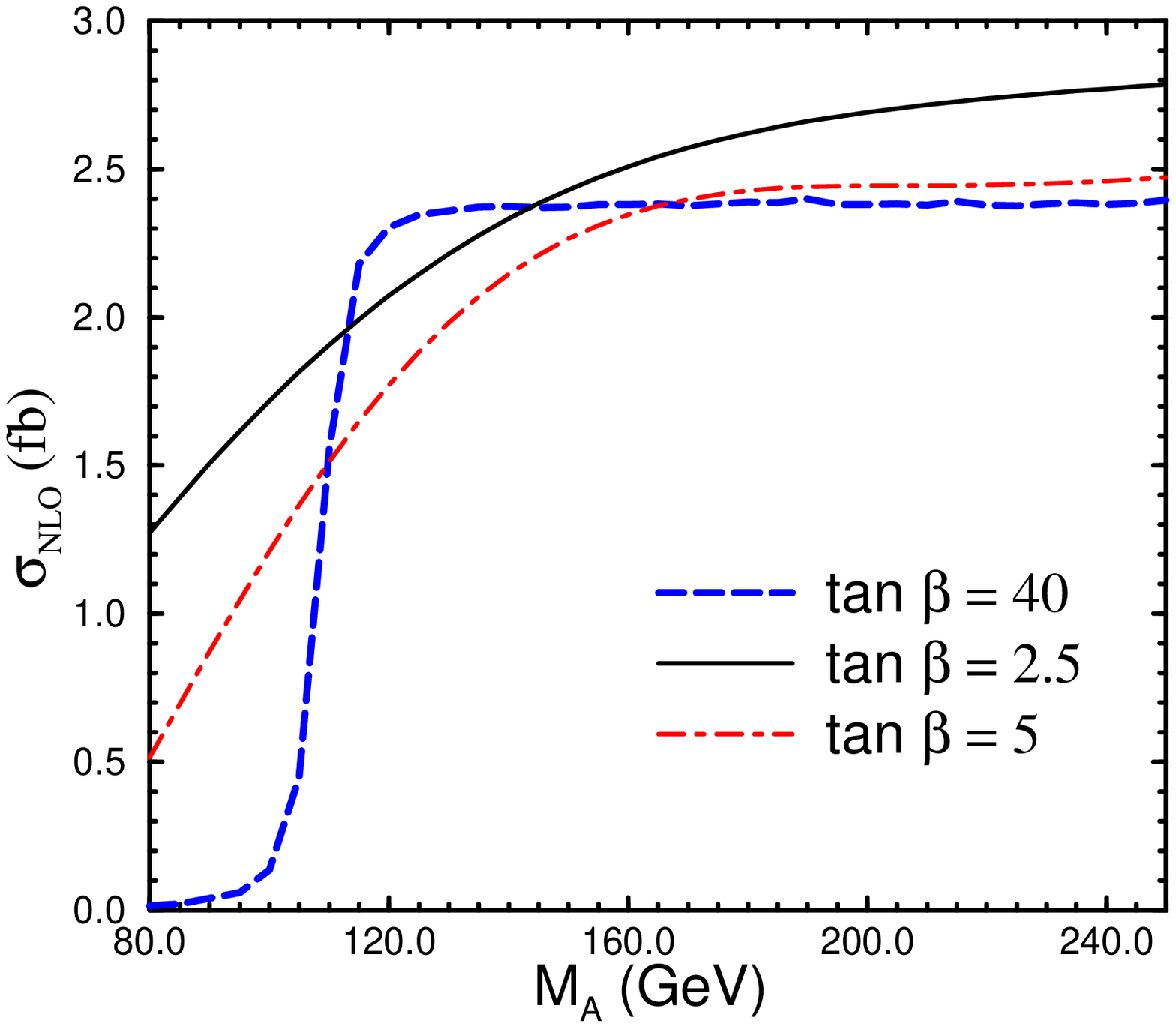}
\caption[]{Next-to-leading order result for 
$e^+e^-\rightarrow t {\overline t} h^0$ at
$\sqrt{s}\!=\!1$~TeV using $K\!=\!0.94$. 
The squarks are taken to have a common mass, $M_S=500~GeV$. }
\label{fig:tevhfig}
\end{figure}

\begin{figure}[t,b]
\centering
\epsfxsize=4.in
\leavevmode\epsffile{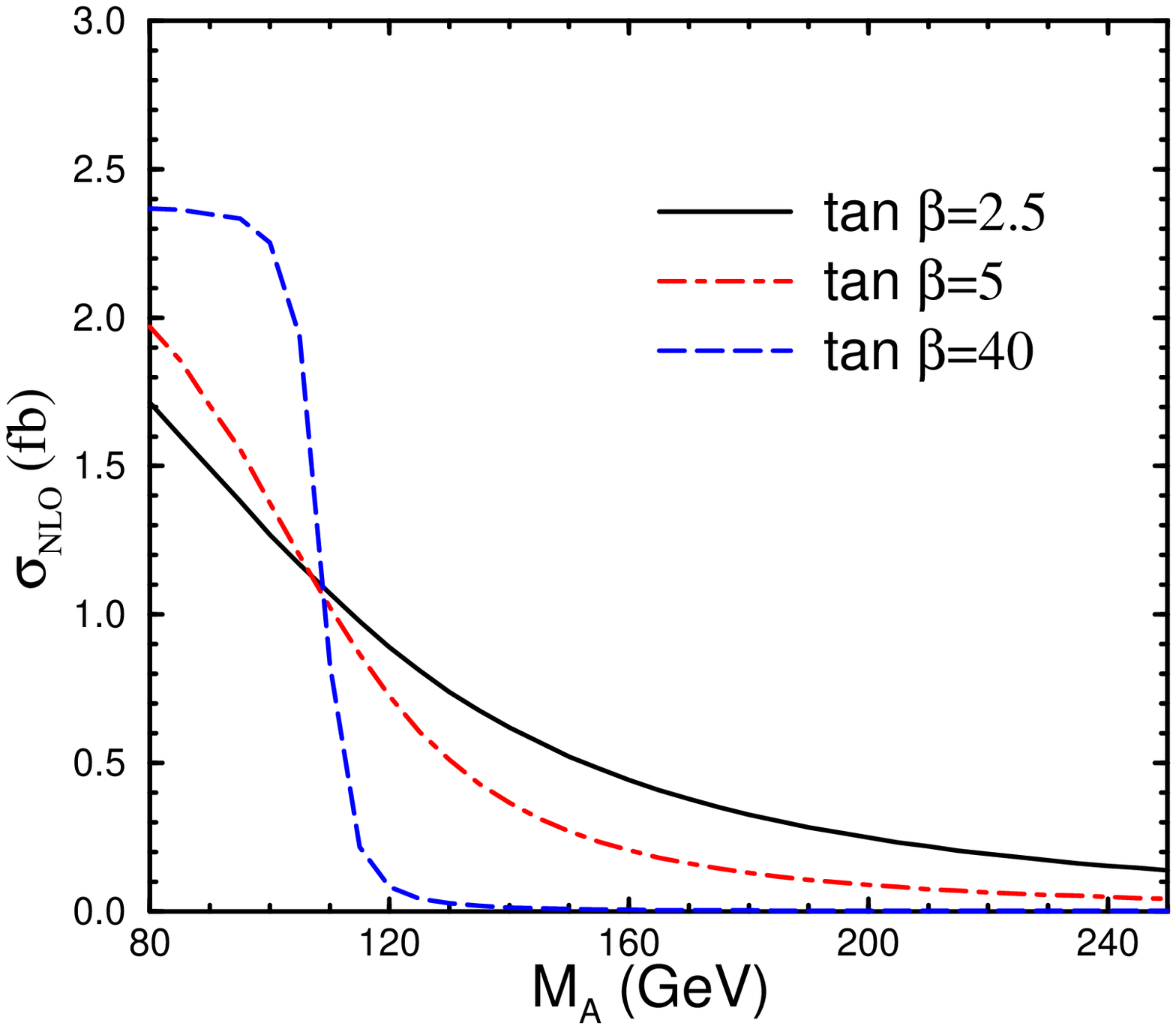}
\caption[]{Next-to-leading order result for 
$e^+e^-\rightarrow t {\overline t} H^0$ at
$\sqrt{s}\!=\!1$~TeV using $K\!=\!0.94$.
The squarks are taken to have a common mass, $M_S=500~GeV$. }
\label{fig:tevhhfig} 
\end{figure}

At $\sqrt{s}\!=\!1$~TeV, the cross section is no longer
dominated by photon exchange and the $Z$ exchange becomes
important.  The ratio of the cross section computed using
only the photon exchange to the total cross section varies
between $70$ and $80~\%$.  However, the contribution from
the $A^0h_i^0$ production of Fig.~\ref{fig:ahfig} is always
less than $1~\%$.  In Ref. \cite{ditt}, the $K$ factor for
Standard Model Higgs production was computed including the
effects of $Z$ exchange.  Using $\alpha_s
(M_t)\!=\!0.11164$, the result of Ref.  \cite{ditt} is that
the $K$ factor at $\sqrt{s}\!=\!1$~TeV is roughly a
constant, $K(1\,\mbox{TeV})\!\sim\!0.94$.  We will use this result
in our plots.  Since the radiation of a Higgs boson from the
$Z$ boson (the second diagram in Fig.~\ref{fig:lofeyndiag})
is small, the $K$ factor is again approximately independent
of the SUSY parameters and depends only on the mass of the
produced Higgs boson.  At high energy where $\sqrt{s}$
is not close to the kinematic threshold, the dominant
contribution to the QCD corrections is from the Higgs vertex
correction computed in Ref. \cite{dr1}.  In this limit the
$K$ factor is roughly,

\beq
K\sim 1-{3\alpha_s(\mu)\over\pi}\sim 0.9\,\,\,,
\eeq

\noindent  in agreement with the
result found from the complete calculation for
$\sqrt{s}\!=\!1$~TeV.\cite{dr2,ditt}
\noindent
Figs.~\ref{fig:tevhfig} and \ref{fig:tevhhfig} show the
rates for $e^+e^-\rightarrow t {\overline t} h_i^0$ at
$\sqrt{s}\!=\!1$~TeV.  The cross section for either
$e^+e^-\rightarrow t {\overline t} h^0$ or $e^+e^-\rightarrow
t {\overline t} H^0$ is greater than $2$~fb throughout most of
the $M_A-\tan\beta$ plane.  For $M_A\!>\!120$~GeV, the rate
for $e^+e^-\rightarrow t {\overline t} h^0$ is relatively
insensitive to $\tan\beta$ and is always between $2.4$ and
$2.8$~fb.  The rate for $e^+e^-\rightarrow t {\overline t}H^0$
is always less than $1$~fb for $M_A\!>\!120$~GeV.  Projected
luminosities for a $1$~TeV machine are around
$200\,\,\mbox{fb}^{-1}$, which would give more than $400$
 $t {\overline t} h^0$ events/year.
 
\begin{figure}[t,b]
\centering
\epsfxsize=4.in 
\leavevmode\epsffile{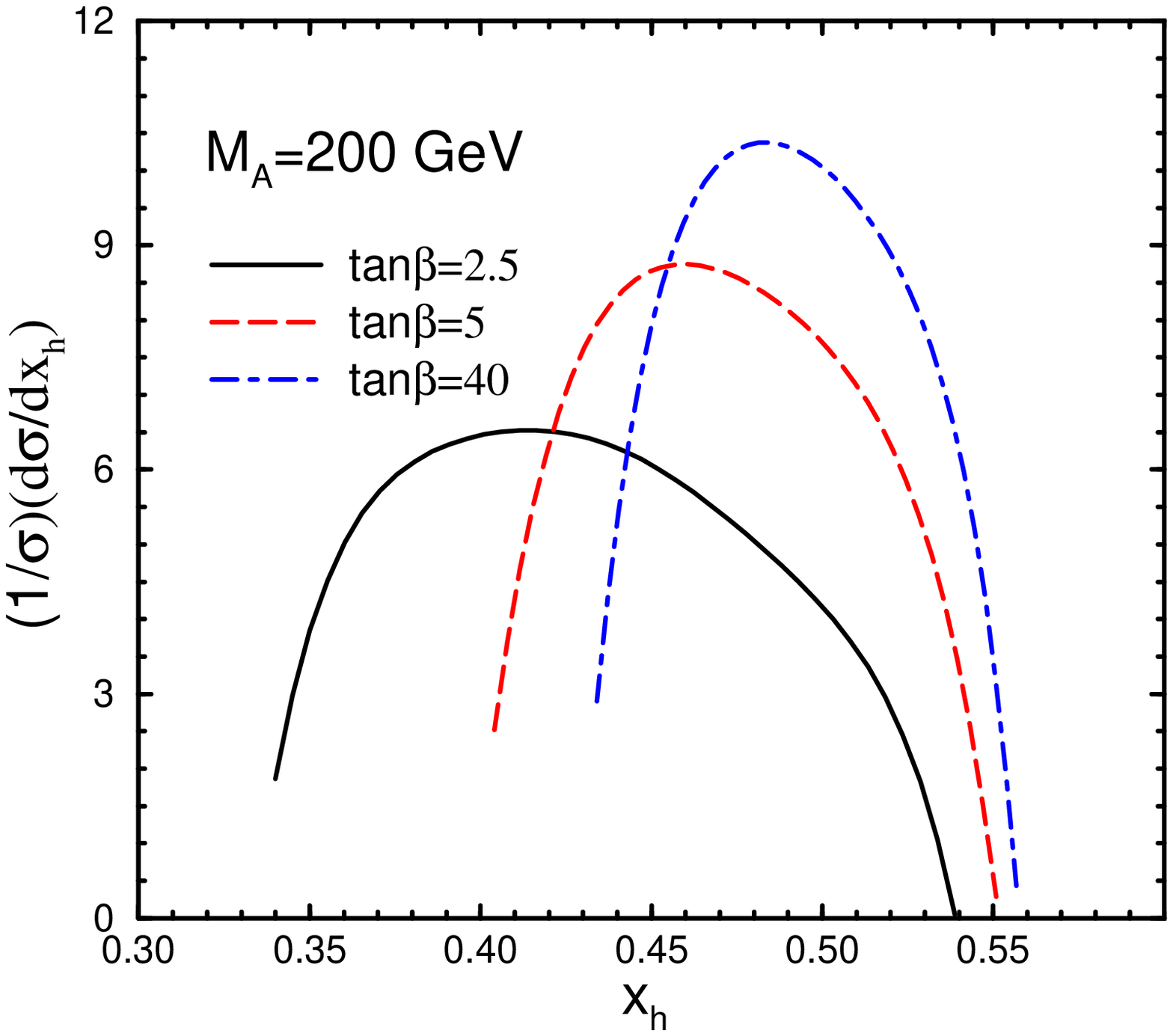}
\caption[]{Dependence of the process $e^+e^-\rightarrow t {\overline t}
  h^0$ at $\sqrt{s}=500~GeV$
on the energy of the Higgs boson,
  $x_h\!=\!2E_h/\sqrt{s}$, for $M_A\!=\!200$~GeV.
The squarks are assumed to have a common mass,
  $M_S\!=\!500$~GeV and the scalar mixing parameters are set
  to zero.}
\label{fig:xhdepfig}                           
\end{figure}

The dependence of the process $e^+e^-\rightarrow t
{\overline t}h^0$ on the energy of the Higgs boson is shown in
Fig.~\ref{fig:xhdepfig}.  As $\tan\beta$ becomes large, the
distribution is peaked at larger values of $x_h\!=\!2
E_h/\sqrt{s}$.
This is a kinematic effect due to the changing of the Higgs boson
mass, since as $\tan\beta$ is increased, the Higgs mass increases
with increasing $M_A$.   
 
\section{$e^+e^-\rightarrow t {\overline t} A^0$}
\label{sec:tta}

%\begin{figure}[t,b]
%\centering
%\epsfxsize=3.in
%\leavevmode\epsffile{tth3.eps}
%\end{figure} 
\begin{figure}[t,b]
\centering
\epsfxsize=4.in
\leavevmode\epsffile{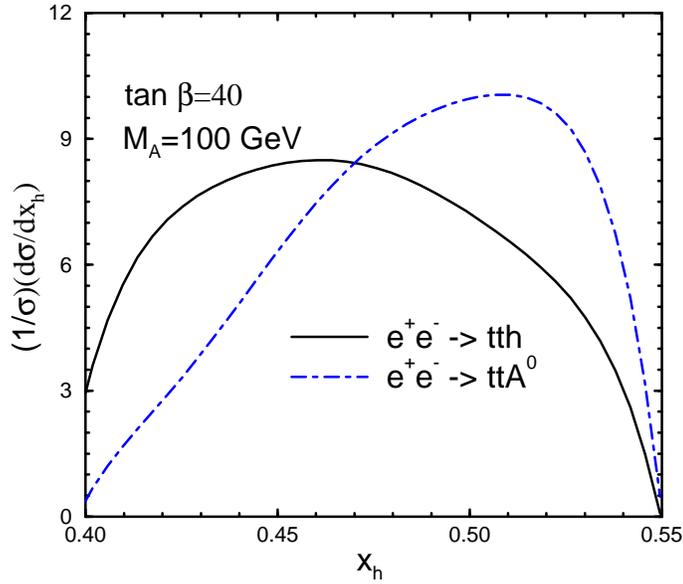}
\caption[]{Distribution in scalar energy for the pseudoscalar,
  $A^0$ and the lightest neutral scalar, $h^0$ of the MSSM at
$\sqrt{s}=500~GeV$.  
The squarks are taken to have a common mass, $M_S=500~GeV$. }
\label{fig:cpdis}
\end{figure} 

The production of a pseudoscalar in association with a heavy
quark pair proceeds through $\gamma$ and $Z$ exchange, and
also through the subprocesses, $e^+e^-\rightarrow A^0
h_i^0\rightarrow A^0 Q {\overline Q}$, where
$h_i^0\!=\!h^0,H^0$.  The rate is given by,\cite{zer1}

\bea 
\frac{d\sigma(e^+e^-\rightarrow Q\bar Q A^0)}{dx_A} &=& 
N_c\frac{\sigma_0}{(4\pi)^2}\left\{
\left[Q_e^2Q_Q^2+\frac{2Q_eQ_Qv_ev_Q}{1-M_Z^2/s}+
\frac{(v_e^2+a_e^2)(v_Q^2+a_Q^2)}{(1-M_Z^2/s)^2}\right]F_1\right.
\nonumber\\
&&\left.+ \frac{v_e^2+a_e^2}{(1-M_Z^2/s)^2}\biggl({1\over 16 c_W^2
s_W^2}\biggr)\left[(2I_{3L}^Q)^2F_2 +F_3+2I_{3L}^Q
F_4 
\right]
\right\}
\,\,\,\,,
\label{dsig0A}
\eea

\noindent 
where $x_A=2E_A/\sqrt{s}$.  Expressions for the coefficients, $F_i(x_A)$, are
 given in Appendix~\ref{app:fcoeff}.

The cross section for $e^+e^-\rightarrow t {\overline t}A^0$ is less
than $10^{-2}$~fb at $\sqrt{s}\!=\!500$~GeV for all values of $\tan\beta$
and $M_A$ in the MSSM and so almost certainly will not be observable.
However, the production of a scalar or of a pseudoscalar in
association with a $t {\overline t}$ pair has been shown to be a
useful mechanism for determining the CP coupling of the Higgs
boson.\cite{cp} In Fig.~\ref{fig:cpdis}, we show the normalized
distribution of the Higgs boson energy for the lightest scalar of the
MSSM and for the pseudoscalar of the MSSM.\footnote{These plots
  include contributions from the $e^+e^-\rightarrow h^0A^0$ resonance,
  which is specific to the MSSM.}  The shape of the distributions is
quite different, suggesting that this may be a useful method for
discriminating between scalar and pseudoscalar Higgs- fermion
couplings.  For this to be practical, of course, requires a model
where the pseudoscalar-fermion coupling is much larger than in the
MSSM.

\section{$e^+e^-\rightarrow b {\overline b} h^0$
          and $e^+e^-\rightarrow b {\overline b} H^0$}
\label{sec:bbh}

\begin{figure}[t,b]
\centering
\epsfxsize=4.in
\leavevmode\epsffile{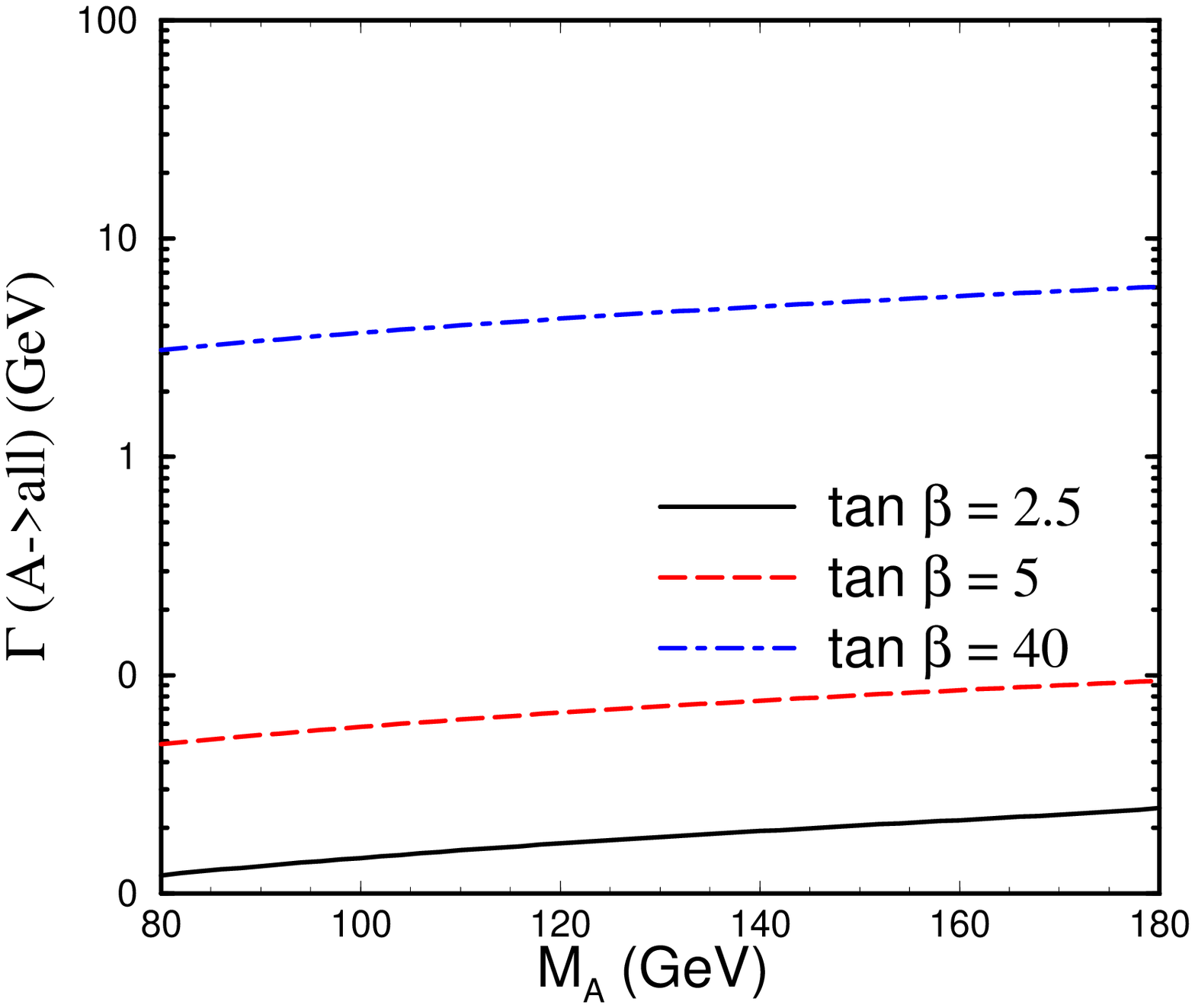}
\caption[]{Total width of the pseudoscalar, $A^0$, in the MSSM including
  NLO QCD corrections and assuming that there are no open
  channels for the $A^0$ to decay into SUSY particles.
The squarks are assumed to have a common mass,
  $M_S\!=\!500$~GeV and the scalar mixing parameters are set
  to zero.}
\label{fig:awidfig}
\end{figure} 

\begin{figure}[t,b]
\centering
\epsfxsize=4.in
\leavevmode\epsffile{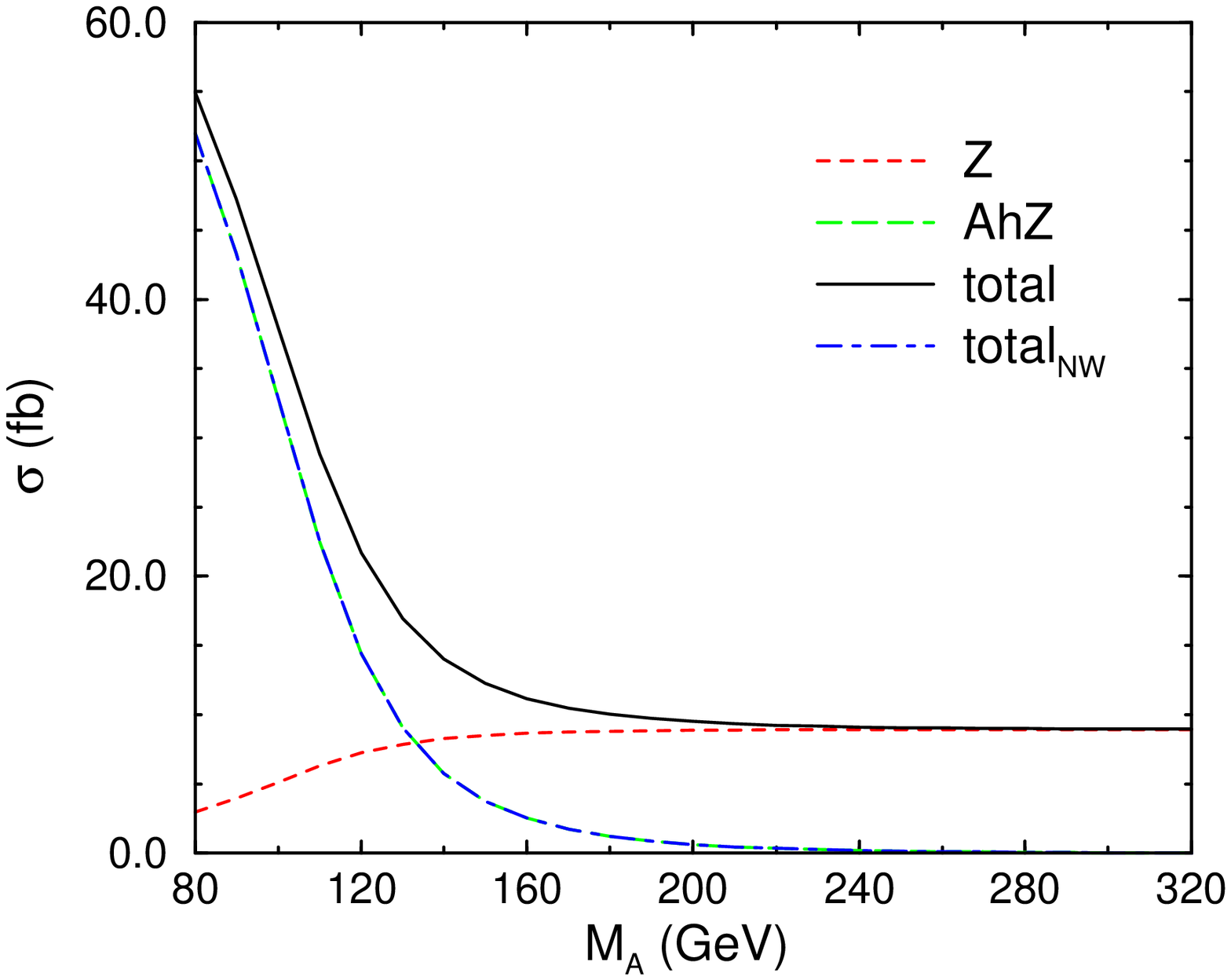}
\caption[]{Contributions to $e^+e^-\rightarrow b {\overline b}h^0$
  at $\sqrt{s}\!=\!500$~GeV, for $\tan\beta\!=\!5$.  The
  curve labelled `NW' is the narrow width approximation of
  Eq.~(\ref{nwsig}) and it includes QCD corrections in the
  resonance region.
The squarks are assumed to have a common mass,
  $M_S\!=\!500$~GeV and the scalar mixing parameters are set
  to zero.}
\label{fig:bb_lh_tb5}
\end{figure} 
  
\begin{figure}[t,b]
\centering
\epsfxsize=4.in
\leavevmode\epsffile{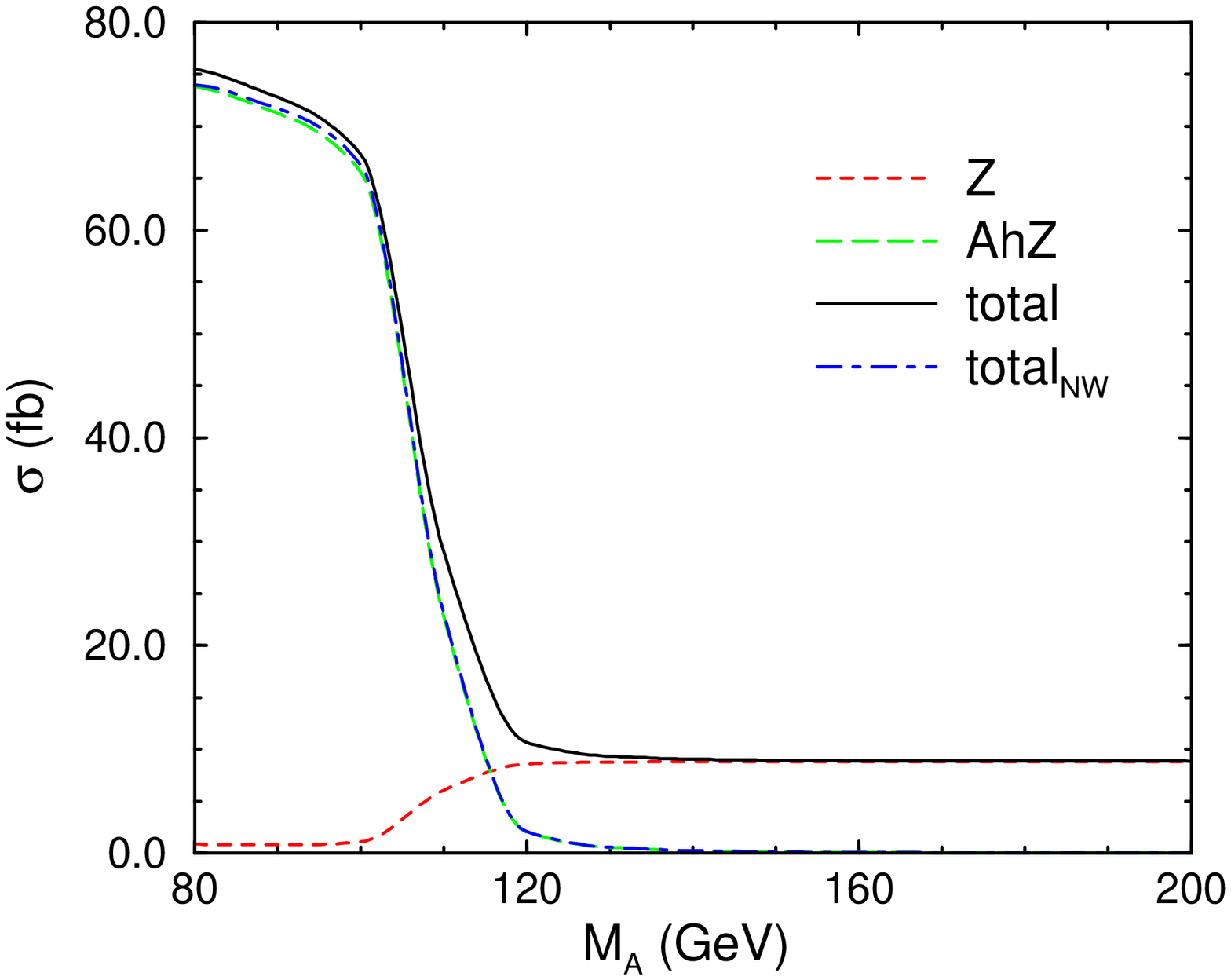}
\caption[]{Contributions to $e^+e^-\rightarrow b {\overline b}h^0$
  at $\sqrt{s}\!=\!500$~GeV, for $\tan\beta\!=\!40$.  The
  curve labelled `NW' is the narrow width approximation of
  Eq.~(\ref{nwsig}) and it includes QCD corrections in the
  resonance region.
The squarks are assumed to have a common mass,
  $M_S\!=\!500$~GeV and the scalar mixing parameters are set
  to zero.}
\label{fig:bb_lh_tb40}
\end{figure}
         
In the MSSM, the Higgs boson can also be produced in
association with $b {\overline b} $ pairs and the
differential cross section is given by Eq.~(\ref{dsig0}).
The physics is significantly different from the production
with a $t {\overline t}$ pair, however.  In the case of the
$b$ quark, the contribution of Fig.~\ref{fig:ahfig} receives
a large resonant enhancement from the decay, $A^0\rightarrow
b {\overline b}$.  This enhancement occurs when $M_{h_i}\sim
M_A$ and $x_h\sim 1$
and so is relevant for $M_A$ below about $120$~GeV for
$e^+e^-\rightarrow b {\overline b} h^0$.

We can estimate the resonant contribution in the narrow
width approximation,

\beq
\sigma(e^+e^-\rightarrow b {\overline b}h_i^0)\mid_{NW}\sim
{3g_{bbh_i}^2 g_{ZAh_i}^2\sigma_0\hat\beta\over 128 \pi
 c_W^2s_W^2}{v_e^2+a_e^2\over (1-
M_Z^2/s)^2}\biggl({M_A\over \Gamma_A}\biggr)
\biggl(3{M_{h_i}^2\over s}+{M_A^2\over s}-1\biggr)\,\,\,,
\label{nwsig}
\eeq 

\noindent
where $\hat\beta$ is given by Eq.(\ref{betadef}) in
Appendix~\ref{app:fcoeff} and $\Gamma_A$ is the total decay
width of the pseudoscalar $A^0$, shown in
Fig.~\ref{fig:awidfig}.  For the parameters considered here,
the pseudoscalar is a very narrow resonance.  This process
is sensitive to the couplings of the $Z$ to the $A^0h_i^0$
pair in the MSSM,\footnote{ The $g_{ZAh_i}$ couplings are
  normalized to the $ZZh_{SM}$ coupling.}

\beqn
g_{ZAh}&=&\cos(\beta-\alpha)\,\,\,,\nonumber \\
g_{ZAH}&=&-\sin(\beta-\alpha)\,\,\,, \nonumber \\
\label{zahdef} 
\eeqn
along with the $b {\overline b}h_i^0$ Yukawa couplings.

Figs.~\ref{fig:bb_lh_tb5} and \ref{fig:bb_lh_tb40} show the different
contributions to the process $e^+e^-\rightarrow {b \overline b} h^0$ for a low
and a high value of $\tan\beta$, at $\sqrt{s}\!=\!500$~GeV \footnote{We do not
consider values of $\tan\beta$ smaller than 5 since the values of $M_h$
corresponding to the resonance region of the $bbh^0$ production mode would be
already experimentally excluded.}. The curve labelled ``$NW$'' is the narrow
width approximation of Eq.~(\ref{nwsig}), labelled as ``\emph{total}'', while
the curve labelled ``$AhZ$'' is only the contribution from the square of the
diagram of Fig.~\ref{fig:ahfig}.  At small $M_A$ ($<120$~GeV), the narrow
width approximation is an excellent approximation to the total rate for these
values of $\tan\beta$.  For smaller $\tan\beta$, the narrow width
approximation becomes increasingly inaccurate, since the $Z$ exchange
contribution becomes more and more relevant. Also at large $M_A$, the rate is
given predominantly by the $Z$ boson exchange contribution and is typically
between $5$ and $10$~fb.\footnote{In the Standard Model, the $Z$
resonance is the dominant contribution.}   In Ref.~\cite{zer1}, only the contribution from the
Higgs coupling to the $b$ quark, (the first diagram of
Fig.~\ref{fig:lofeyndiag}), was included in the figures.  This is only valid
for $M_A$ away from the resonance region.  For $M_A$ near the resonance
region, Ref.~\cite{zer1} significantly underestimates the size of this
process.

In the narrow width approximation, the QCD corrections to the rate are
trivially included by including the QCD corrections to the
pseudo-scalar width and this is done in Figs.~\ref{fig:bb_lh_tb5} and
\ref{fig:bb_lh_tb40}.  The QCD ${\cal O}(\alpha_s)$ corrections to
$\Gamma_A$ can be found in Ref.~\cite{spir}.  Away from the resonance,
(large $M_A$), inclusion of the QCD corrections would require a
complete calculation, which we do not include in the present analysis
since the interesting region is near the resonance where the rate is
enhanced.

\begin{figure}[t,b]
\centering
\epsfxsize=4.in
\leavevmode\epsffile{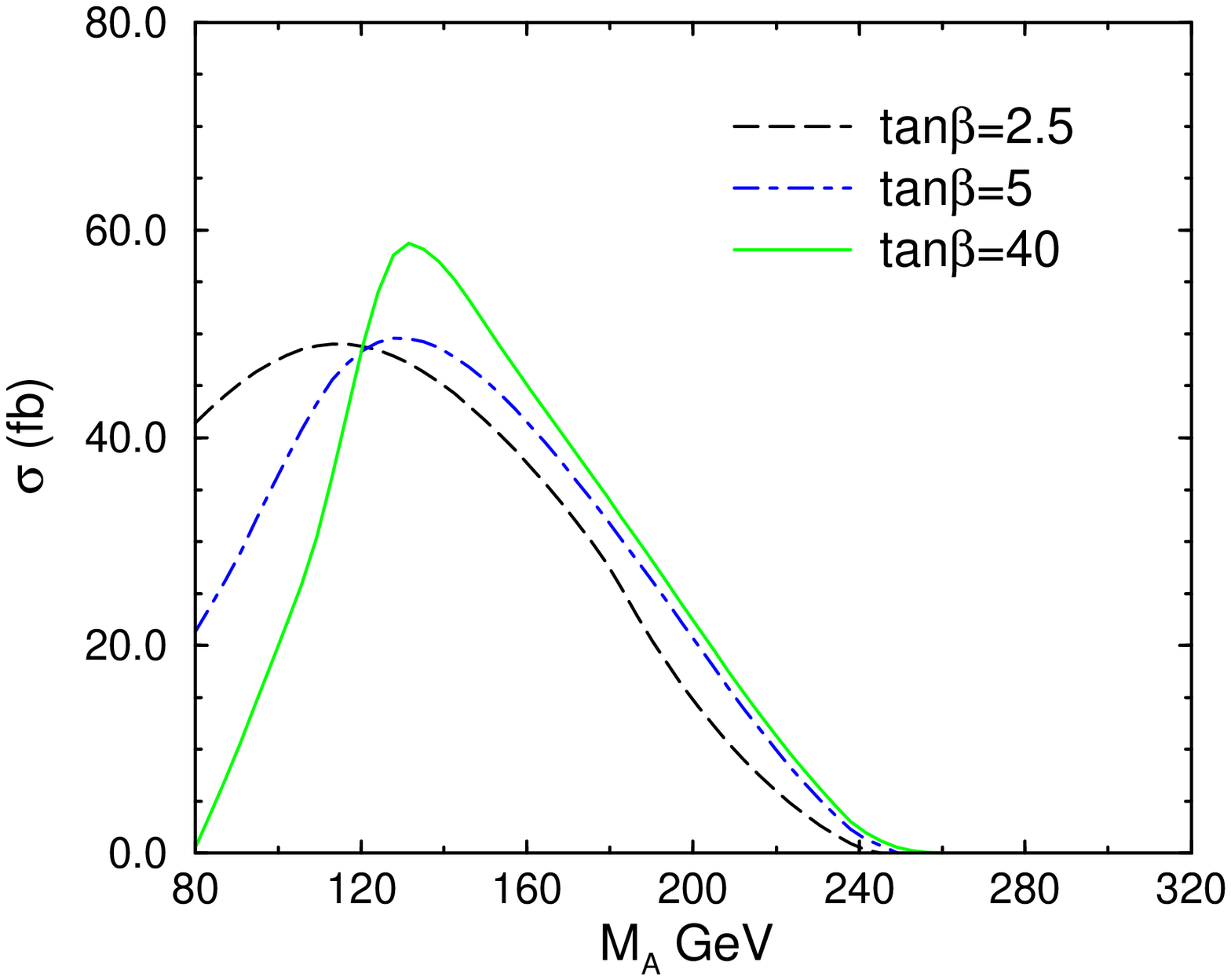}
\caption[]{Cross section for $e^+e^-\rightarrow b {\overline b}H^0$
at $\sqrt{s}\!=\!500$~GeV, including QCD corrections in the
narrow width approximation.
The squarks are assumed to have a common mass,
  $M_S\!=\!500$~GeV and the scalar mixing parameters are set
  to zero.}
\label{fig:bbhhmafig} 
\end{figure}

For the heavy Higgs production, the narrow width approximation is an
excellent approximation for all values of $\tan\beta$ so the QCD
corrections can be accurately included everywhere.  The rate for $b
{\overline b}H^0$ production is shown in Fig.~\ref{fig:bbhhmafig}.  For
$\tan\beta<5$, the cross section is larger than $20$~fb even for
$M_A\!\sim\! 200$~GeV.  For $\tan\beta\!>\!5$, the rate is greater than
$20$~fb for $M_A\!>\!110$~GeV.  This process can potentially be used to
probe the couplings of the heavier neutral Higgs boson and obtain a
precise measurement of the product of the Higgs couplings,
$g_{bbH}g_{ZAH}$.

The same type of
behavior is still present if we vary over the possible values
of the scalar masses and mixing parameters or if we look at
$\sqrt{s}\!=\!1$~TeV.

\section{$e^+e^-\rightarrow b {\overline b} A^0$}
\label{sec:bba}

\begin{figure}[t,b]
\centering
\epsfxsize=4.in
\leavevmode\epsffile{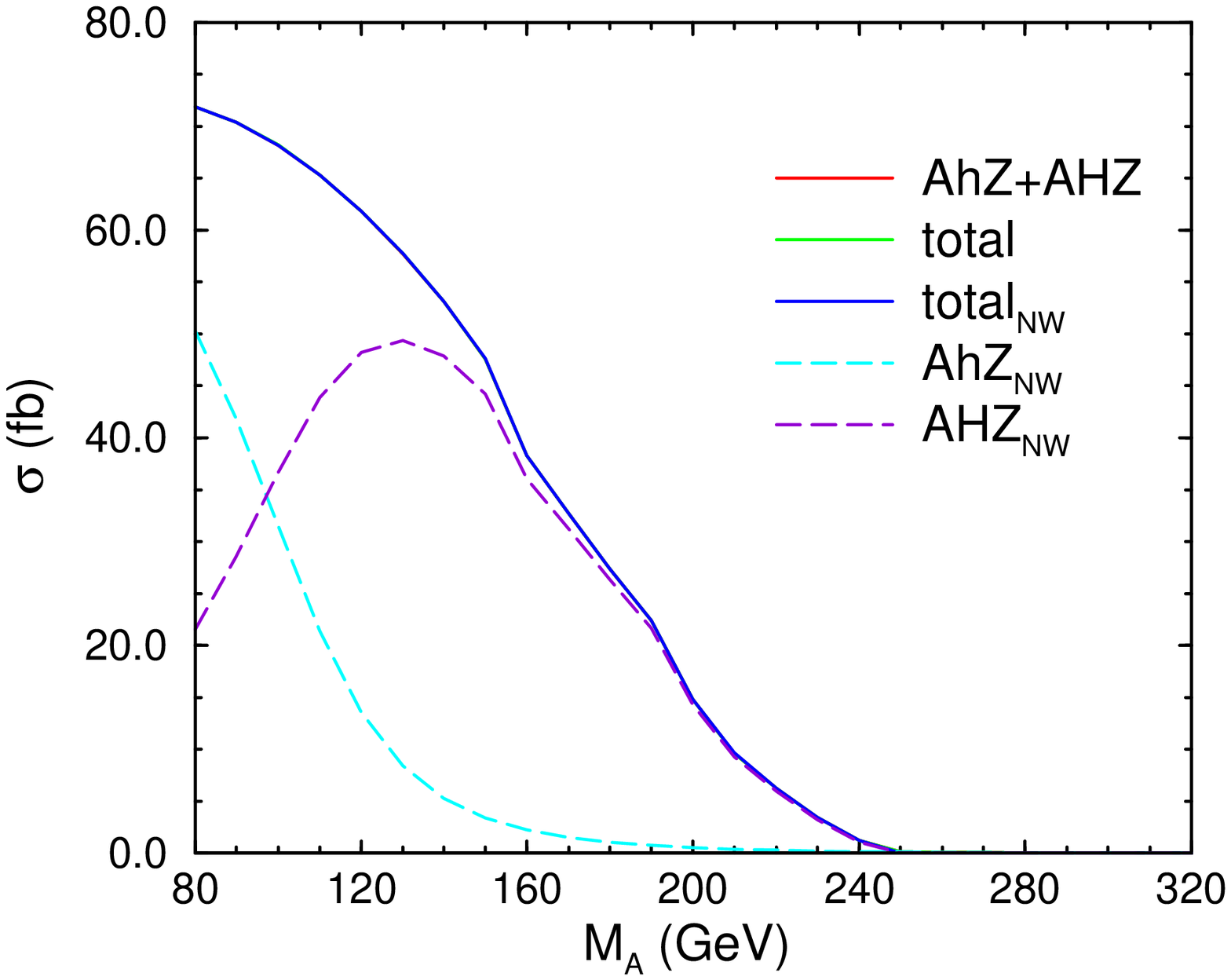}
\caption[]{Total and partial contributions to the
  cross section for $e^+e^-\rightarrow b {\overline b}A$ at
  $\sqrt{s}\!=\!500$~GeV and $\tan\beta\!=\!5$, including
  QCD corrections.
The squarks are assumed to have a common mass,
  $M_S\!=\!500$~GeV and the scalar mixing parameters are set
  to zero.}
\label{fig:bba_tb5}
\end{figure}

\begin{figure}[t,b]
\centering
\epsfxsize=4.in
\leavevmode\epsffile{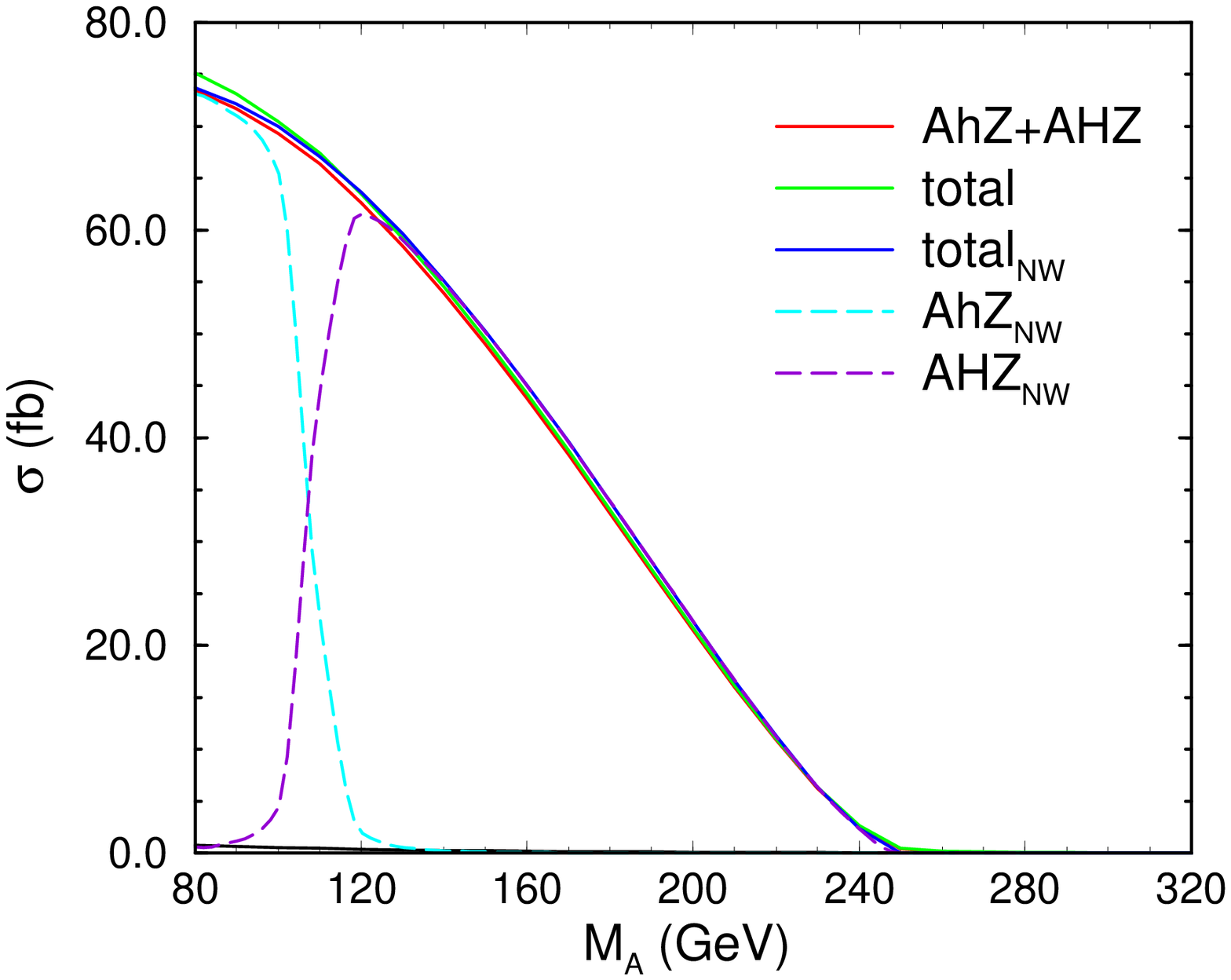}
\caption[]{Total and partial contributions to the
  cross section for $e^+e^-\rightarrow b {\overline b}A$ at
  $\sqrt{s}\!=\!500$~GeV at $\tan\beta\!=\!40$, including
  QCD corrections.
The squarks are assumed to have a common mass,
  $M_S\!=\!500$~GeV and the scalar mixing parameters are set
  to zero.}
\label{fig:bba_tb40}
\end{figure}

We can safely work in the narrow width approximation also
for the case of $e^+e^-\rightarrow b {\overline b} A^0$
production.  In fact, in this case the $h_i^0$ resonances are
completely dominant and the exact cross section can be
distinguished from the one obtained using the narrow width
approximation only at very high values of $\tan\beta$. The
cases of $\tan\beta\!=\!5$ and $\tan\beta\!=\!40$ are
illustrated in Figs.~\ref{fig:bba_tb5} and
\ref{fig:bba_tb40}.  We can see that for $\tan\beta\!=\!5$
the exact cross section is indistinguishable from the one
obtained in the narrow width approximation and the only
relevant contribution is given by $e^+e^-\rightarrow
A^0h_i^0\rightarrow Ab\bar b$.

Unlike $t{\overline t}A^0$ production, the process
$e^+e^-\rightarrow b {\overline b} A^0$ is not suppressed
relative to $e^+e^-\rightarrow b {\overline b} h^0$
production.
\noindent The resonant behavior of the $bbA^0$ production mode is 
very stable with respect to the variation of the scalar mass and
mixing parameters of the MMSM. Different parametrizations only affect
the values of $M_h$ and $M_H$ at resonance.

\section{Conclusion}
\label{sec:concl}

We have considered the production of the Higgs bosons of the
MSSM in association with a heavy quark pair.  The processes
$e^+e^-\rightarrow t {\overline t} h^0 (H^0)$ have rates around
$1$~fb at $\sqrt{s}\!=\!500$~GeV over much of the parameter
space of the MSSM. They can reach rates of $2$~fb in selected
regions of the parameter space at $\sqrt{s}\!=\!500$~GeV and
over most of the parameter space at $\sqrt{s}\!=\!1$~TeV.
Observation of these processes could provide a precise
measurement of the $tth_i^0$ couplings in a region not
accessible at LEP2 and with $50\,\mbox{fb}^{-1}$, the
$tth^0$ process could give a $90~\%~ {\it c.l.}$
measurement with a statistical error of
\beq
{\delta g_{tth}\over g_{tth}}< \pm10\%
\eeq
over much of the $M_A-\tan \beta$ parameter
space.   The shape of the Higgs boson energy
spectrum in these processes is quite sensitive to the value
of $\tan\beta$.  The rate for $e^+e^-\rightarrow t
{\overline t} A^0$ is too small to be observed.

For $M_A\!>\!120$~GeV, the rate for $e^+e^-\rightarrow b
{\overline b} h^0$ is also small, on the order of a few $fb$
at $\sqrt{s}\!=\!500$~GeV.  However, in the resonance
region, $M_A\sim M_{h}$, $b {\overline b}h^0$ production
receives a large enhancement.  This leads to the possibility
of probing
%g_{bbh}g_{ZAh}$
above the threshold accessed by LEP2.  The rates for
$e^+e^-\rightarrow b {\overline b}H^0$ and
$e^+e^-\rightarrow b {\overline b} A^0$are greater than
$20~fb$ throughout much of the MSSM parameter space and
offer a unique window for studying the couplings of the
pseudoscalar and heavier neutral Higgs bosons.

\newpage
\appendix
\section{Coefficients for $e^+e^-\rightarrow Q {\overline Q} h_i^0$
Production}
\label{app:gcoeff}

The cross section for the process $e^+e^-\rightarrow Q {\overline Q} h_i^0$
($h_i^0\!=\!h^0,H^0$) is given in Eq.~(\ref{dsig0}) in terms of coefficients,
$G_i$.  The coefficients $G_1$ and $G_2$ describe the radiation of the Higgs
boson from the heavy quark (both photon and $Z$ boson exchange) and are given
by,\cite{zer1,dr2}\footnote{$G_1,...G_6$ agree with those of Ref. \cite{zer1}.
$G_7$ corrects a sign error in Ref. \cite{zer1}. 
We thank J.~Kalinowski for a discussion of Ref. \cite{zer1}. 
} 

\bea
\label{th_coeff}
G_1&=&\frac{2\,g_{QQh_i}^2}{s^2\left(\hat\beta^2 - x_h^2 \right)x_h}
     \left(\phantom{\frac{1}{2}}\!\!\!\!
        -4\hat\beta\,\left( 4 M_Q^2 - M_{h_i}^2 \right) \,
        \left( 2 M_Q^2 + s \right) x_h + \right.\\
&& \!\!\!\!\!\!\!\!\!\!\!\!\!\!\!\!\!
        \left.\left(\hat\beta^2 - x_h^2 \right) 
        \left( 16M_Q^4 + 2M_{h_i}^4 - 
          2M_{h_i}^2 s x_h + s^2 x_h^2 - 4 M_Q^2  
        \left( 3 M_{h_i}^2 - 2 s - 2 s x_h \right)  \right) \,
        \Lambda
       \right)\,\,\,, \nonumber\\
G_2&=& \frac{-2\,g_{QQh_i}^2}{s^2\,\left(\hat\beta^2 - x_h^2 \right) x_h}
     \left(\phantom{\frac{1}{2}}\!\!\!\! \hat\beta\, x_h
        \left( -96 M_Q^4 + 24 M_Q^2 M_{h_i}^2 - 
          \left( -M_{h_i}^2 + s + s x_h \right) 
           \left( -\hat\beta^2 + x_h^2 \right)  \right)\right.  +
     \nonumber\\
&& \!\!\!\!\!\!\!\!\!\!\!\!\!\!\!\!\!
        \left.   2\left( \hat\beta^2 - x_h^2 \right) 
        \left( 24 M_Q^4 + 2\left( M_{h_i}^4 - M_{h_i}^2 s x_h \right)  + 
          M_Q^2\left( -14 M_{h_i}^2 + 12 s x_h + s x_h^2 \right)  \right) 
        \Lambda
       \right) \,\,\,,\nonumber
\eea

\noindent where we have defined

\beqn
\Lambda&\equiv&\log\left({\frac{x_h+\hat\beta}{x_h-\hat\beta}}\right) 
\nonumber \\
X_i&\equiv &\left( M_{h_i}^2 - M_Z^2 + s - s x_h \right)
\eeqn

\noindent and

\be
{\hat \beta}=\biggl\{{[x_h^2-(x_h^{min})^2][x_h^{max}-x_h]
\over x_h^{max}-x_h+{4 M_Q^2\over s}}\biggr\}^{1/2}\,\,\,\,,
\label{betadef}
\ee 
\noindent with $x_h^{min}=2 M_{h_i}/\sqrt{s}$ and $x_h^{max}=1-4M_Q^2/s
+M_{h_i}^2/s$.

\noindent 
The other four coefficients, $G_3,\ldots,G_6$ describe the
emission of a Higgs boson from the $Z$-boson (plus relative
interference terms) and can be written in the following
form,

\bea
\label{Zh_coeff}
G_3&=& \frac{-2\,\hat\beta g_Z^2 M_Q^2}
{M_Z^2 s^2X_i^2}
     \left( 4 M_{h_i}^4 + 12 M_Z^4 + 2 M_Z^2 s x_h^2 + 
       s^2\left( -1 + x_h \right) x_h^2 - \right.\nonumber\\
   && \left.   M_{h_i}^2\left( 8 M_Z^2 + s\left( -4 + 4 x_h + x_h^2 \right) 
           \right)  \right)\,\,\,, \nonumber\\
G_4 &=& \frac{\hat\beta g_Z^2 M_Z^2}
 {6s^2X_i^2}
     \left( 48 M_Q^2 + 12 M_{h_i}^2 - s\left( -24 + \hat\beta^2 + 24 x_h - 
          3 x_h^2 \right)  \right)\,\,\,, \\
G_5 &=&
\frac{4\,g_{QQh_i}\,g_Z\,M_Q}
{M_Z\,s^2 X_i  }
     \left(\phantom{\frac{1}{2}}\!\!\!\! \hat\beta\,s  \left( 6 M_Z^2 + 
     x_h\left( -M_{h_i}^2 - s + s x_h \right)\right) + \right.\nonumber\\
  && \left.   2\left( M_{h_i}^2\left( M_{h_i}^2-3M_Z^2 + s - s x_h \right)  + 
          M_Q^2\left( -4 M_{h_i}^2 + 12 M_Z^2 + s x_h^2 \right)  \right) \,
        \Lambda
       \right)\,\,\,, \nonumber\\
G_6 &=&
\frac{-8\,g_{QQh_i} \,g_Z\,M_Q\,M_Z}
{s^2 X_i }
     \left( \hat\beta\,s + \left( 4 M_Q^2 - M_{h_i}^2 + 2 s - s x_h \right) 
        \Lambda
       \right) \nonumber\,\,\,\,,
\eea

\noindent  
where $g_Z$ denotes the coupling of the Higgs boson to the $Z$ boson,

\be 
g_Z\!\equiv\! (\sqrt{2}G_F)^{1/2}\,M_Z g_{ZZh_i}\,\,\,,
\ee 

\noindent where
\beqn
g_{ZZh}&=& sin(\beta-\alpha)\,\,\,,\nonumber \\
g_{ZZH}&=& cos(\beta-\alpha)\,\,\,. 
\eeqn

\noindent 
The contribution from $A^0h_i^0$ production (see Fig.~\ref{fig:ahfig}) is
given by

\beqn 
G_7&=&-{ g_{QQA} g_{ZAh_i}\over Y_i} \biggl\{ 2 \beta (4
M_{h_i}^2-sx_h)\biggl[- g_{ZAh_i}g_{QQA} {sx_h-s-M_{h_i}^2\over Y_i}
-{4 M_Q I_{3L}^Q g_{ZZh_i}\over M_Z}\biggr]
\nonumber \\
&&+\frac{4 I_{3L}^Qg_{QQh_i}}{s} \biggl[ 2sx_h \beta
(sx_h-s-M_{h_i}^2)-4\biggl(M_{h_i}^2(sx_h-s-M_{h_i}^2) \nonumber \\ &&
+ M_Q^2(4M_{h_i}^2-sx_h^2)\biggr)\Lambda\biggr]\biggr\}\,\,\,, 
\eeqn

\noindent
where we have defined 

\beq
Y_i = M_{h_i}^2-M_A^2+s-sx_h\,\,\,\,,
\eeq

\noindent 
while $g_{QQA}$ and $g_{ZAh_i}$ are given in Eqs.~(\ref{yukdef})
(\ref{zahdef}).

\section{Coefficients for $e^+e^-\rightarrow Q {\overline Q}A^0$}
\label{app:fcoeff}

The coefficients of Eq.~(\ref{dsig0A}) are\footnote{Our result for
  $F_4$ ($F_3$) is a factor of 2
$(4)$ larger than the result of Ref. \cite{zer1}.}
            
\beqn
F_1&=& 2g_{QQA}^2\biggl\{\biggl[x_A-{2 M_A^2\over s^2 x_A}\biggl(
sx_A-M_A^2+2M_Q^2\biggr)\biggr]\Lambda +{4\frac{M_A^2}{s^2} 
(s+2M_Q^2){\hat \beta \over {{\hat \beta}}^2-x_A^2}}\biggr\} \,\,\,,
\nonumber \\
F_2 &=& 2g_{QQA}\left\{\hat\beta(1+x_A)-
\frac{2}{s^2x_A}\left[2M_A^4+M_Q^2sx_A^2-2M_A^2(M_Q^2+sx_A)\right]\Lambda
\right.\nonumber\\
&+&\left.
\frac{M_A^2}{s^2}\frac{\hat\beta}{\hat\beta^2-x_A^2}\left(
-24M_Q^2+\hat\beta^2 s-sx_A^2\right)\right\}\,\,\,,
\\
F_3&=&
2\hat\beta\biggl[\sum_i{g_{QQh_i}g_{ZAh_i}\over Z_i}\biggl]^2
\biggl(sx_A-s-M_A^2+4M_Q^2\biggr)\biggl(4M_A^2-sx_A^2\biggr) \,\,\,,
\nonumber 
\\
F_4 &=&  \frac{4}{s}
\sum_i\left[\frac{g_{QQA}g_{QQh_i}g_{ZAh_i}}{Z_i}\right]
\left\{\hat\beta sx_A(M_A^2+s-sx_A)\right.
\nonumber\\
&-&\left.
2\left[M_A^4+M_Q^2sx_A^2+M_A^2(-4M_Q^2+s-sx_A)\right]\Lambda\right\}
\,\,\,,
\nonumber
\eeqn 

\noindent where we define
\be
Z_i = M_A^2-M_{h_i}^2+s-sx_A\,\,\,\,.
\ee
  
\end{document}